%% file: ms_astroph.tex
\documentclass[12pt,preprint]{aastex}

\newcommand{\teff}{$T_{\rm eff}$}
\newcommand{\tc}{$T_{\mathrm{c}}$}
\newcommand{\swan}{C$_{2}$}

\begin{document}

\title{ABUNDANCES OF STARS WITH PLANETS: TRENDS WITH CONDENSATION TEMPERATURE\altaffilmark{1,2}}

\altaffiltext{1}{Based on observations with the High Resolution Spectrograph on the
                 Hobby-Eberly Telescope, which is operated by McDonald Observatory on behalf
		 of the University of Texas at Austin, Pennsylvania State University,
		 Standford University, the Ludwig-Maximilians-Universit{\"a}t M{\"u}nchen,
		 and the Georg-August-Universit{\"a}t, G{\"o}ttingen.}
\altaffiltext{2}{Based on observations made with the FEROS instrument on the MPG/ESO 2.2-m
                 telescope at La Silla (Chile), under the agreement ESO-Observat{\'o}rio 
		 Nacional/MCT.}

\author{Simon C. Schuler\altaffilmark{3,4}, Davin Flateau\altaffilmark{5}, Katia 
        Cunha\altaffilmark{3,6,7}, Jeremy R. King\altaffilmark{8}, Luan Ghezzi\altaffilmark{6}, 
	AND Verne V. Smith\altaffilmark{3}}
\affil{
   \altaffiltext{3}{National Optical Astronomy Observatory, 950 North Cherry
   Avenue, Tucson, AZ, 85719  USA; sschuler@noao.edu, kcunha@noao.edu, vsmith@noao.edu}
   \altaffiltext{4}{Leo Goldberg Fellow}
   \altaffiltext{5}{Department of Physics, University of Cincinnati, Cincinnati, OH
   45221  USA; flateadc@mail.uc.edu}
   \altaffiltext{6}{Observat{\'o}rio Nacional, Rua General Jos{\'e} Cristino, 77, 20921-400,
   S{\~a}o Crist{\'o}v{\~a}o, Rio de Janeiro, RJ, Brasil; luan@on.br}
   \altaffiltext{7}{Steward Observatory, University of Arizona, 933 North Cherry Avenue, 
   Tucson, AZ 85721  USA}
   \altaffiltext{8}{Department of Physics and Astronomy, Clemson University, 118
   Kinard Laboratory, Clemson, SC, 29634  USA; jking2@ces.clemson.edu}
   }

\begin{abstract}
Precise abundances of 18 elements have been derived for ten stars known to host giant planets from
high signal-to-noise ratio, high-resolution echelle spectroscopy.  Internal uncertainties in the
derived abundances are typically $\lesssim 0.05$ dex.  The stars in our sample have all been 
previously shown to have abundances that correlate with the condensation temperature (\tc) of the 
elements in the sense of increasing abundances with increasing \tc; these trends have been interpreted 
as evidence that the stars may have accreted H-depleted planetary material.  Our newly derived 
abundances also correlate positively with \tc, although slopes of linear least-square fits to the 
[m/H]-\tc\ relations for all but two stars are smaller here than in previous studies.  When 
considering the refractory elements (\tc\ $> 900$ K) only, which may be more sensitive to planet 
formation processes, the sample can be separated into a group with positive slopes (four stars) and 
a group with flat or negative slopes (six stars).  The four stars with positive slopes have very 
close-in giant planets (three at 0.05 AU) and slopes that fall above the general Galactic chemical 
evolution trend.  We suggest that these stars have accreted refractory-rich planet material but not 
to the extent that would increase significantly the overall stellar metallicity.  The flat or negative 
slopes of the remaining six stars are consistent with recent suggestions of a planet formation 
signature, although we show that the trends may be the result of Galactic chemical evolution.
\end{abstract}


\section{INTRODUCTION}
\label{s:intro}
The primary objective of chemical abundance studies of planetary host stars is to identify possible
vestiges of the planet formation process that will lead to a greater understanding of how planets 
form and evolve.  The validity of this endeavor was verified shortly after the discovery of the 
first planet orbiting a solar-type star \citep{1995Natur.378..355M} when 
\citet{1997MNRAS.285..403G,1998A&A...334..221G} found that host stars, in general, have larger Fe 
abundances than stars without known planets.  The metal-rich nature of stars with giant planets 
has been confirmed by various groups 
\citep[e.g.,][]{2001A&A...373.1019S,2005ApJ...622.1102F,2010ApJ...720.1290G}, and substantial
observational evidence indicates that it is an intrinsic property of these planetary systems 
\citep[e.g.,][]{2005ApJ...622.1102F,2010ApJ...725..721G}.  Core-accretion models of planet
formation \citep[e.g.,][]{2004ApJ...616..567I} naturally account for this giant planet-metallicity 
correlation.

An alternative explanation for the enhanced metallicities of stars with giant planets was proposed
by \citet{1997MNRAS.285..403G}.  He suggested that the metallicities of the host stars are 
not primordial but are the result of self-enrichment, i.e., the accretion of H-depleted material 
onto the star as a result of dynamical processes in the protoplanetary disk.  Gonzalez postulated 
that if stars with planets accrete fractionated disk material, their photospheric abundances 
should correlate with the condensation temperatures (\tc) of the elements such that high-\tc\ 
refractory elements are more abundant than low-\tc\ volatile elements.  Whereas the infall of 
planetary debris onto host stars may be a common occurrence in planet forming disks 
\citep[for a review, see][]{2008ApJ...685.1210L}, it is unclear from modeling efforts if accreted 
material would leave an observable imprint on the a stellar photosphere 
\citep{2001ApJ...556L..59P,2002ApJ...566..442M,2004ApJ...605..874V}.  Attempts to 
identify trends with \tc\ \citep{2001AJ....121.3207S,2006A&A...449..809E,2006MNRAS.367L..37G} 
resulted in finding no significant differences between stars with and without giant planets, 
although \citet[][henceforth S01]{2001AJ....121.3207S} and 
\citet[][henceforth E06]{2006A&A...449..809E} noted that small subsets of stars with planets stood 
out as having particularly strong correlations of increasing abundances with increasing \tc.  
Furthermore, S01 found that the stars with the strong correlations have planets that are 
on much closer orbits, on average, than stars not showing the possible abundance trend.

\citet[][henceforth M09]{2009ApJ...704L..66M} revisited the idea that accretion of disk material, 
while maybe not the primary mechanism responsible for the observed enhanced metallicities, may 
imprint \tc\ trends in the photospheres of planet host stars, with results that are contrary to 
original expectations.  They showed that the Sun is {\it deficient} in refractory elements relative to 
volatile elements when compared to the mean abundances of 11 solar twins (stars with stellar 
parameters that are nearly identical to those of the Sun) and that the abundance differences 
correlate strongly with \tc\ in the sense that the abundances decrease with increasing \tc.  
This trend is interpreted by the authors as a possible signature of terrestrial planet formation 
in the Solar System, suggesting that the refractory elements depleted in the solar photosphere are 
locked up in the terrestrial planets.  

In a comparison of solar refractory abundances to the refractory abundances of solar twins and solar 
analogs (stars with stellar parameters similar to those of the Sun) it was found that $\sim 85\%$ of 
the stars do not show the putative terrestrial planet signature, i.e., they are enhanced in refractory 
elements relative to the Sun \citep{2009A&A...508L..17R,2010A&A...521A..33R}.  These studies speculate 
that the remaining $\sim15\%$ of the stars, which have abundance patterns similar to the Sun, are 
terrestrial planet hosts.  Subsequently, \citet{2010MNRAS.407..314G} investigated the abundances of 
refractory elements versus \tc\ trends for a sample of stars with and without known giant planets.  
Stars with giant planets were found to have more negative trends (decreasing abundances with 
increasing \tc) than stars without known planets; moreover, the most metal-rich stars with giant 
planets have the most negative trends.  These results potentially indicate that the depleted 
abundances of refractory elements in stellar photospheres are a consequence of both terrestrial and 
giant planet formation \citep{2010MNRAS.407..314G}.  Recently, \citet{2010ApJ...720.1592G} studied a 
sample of solar twins and analogs with and without planets, and found similar abundance patterns for 
each sample, including two stars with terrestrial super-Earth-type planets; they have suggested that 
the abundance pattern identified by M09 may not be related to planet formation.  
\citet{2010A&A...521A..33R} in turn pointed out that the analysis of \citet{2010ApJ...720.1592G} 
included both volatile and refractory elements, and that the planet signature is more robust among 
the refractories.  In a reanalysis of the Gonz{\'a}lez Hern{\'a}ndez et al. data for the two stars 
with super-Earth-type planets, \citet{2010A&A...521A..33R} find abundance patterns consistent with 
the planet signature.

Here we present precise abundances of 18 elements for ten stars with known giant planets derived 
homogeneously from high-quality, high-resolution echelle spectroscopy.  The target stars were
taken from the aforementioned works of S01 and E06, and are a subset of those that were reported 
to have the strongest correlations of increasing abundances with increasing \tc.  Thus, according 
to these studies, the stars are candidates for having accreted fractionated refractory-rich 
material.  We compare our high-precision abundances to \tc, for all elements and for refractory 
elements (\tc\ $> 900$ K) only, to further investigate possible planet formation signatures.

\section{OBSERVATIONS AND DATA REDUCTION}
\label{s:obs}
The stars studied here are distributed at both northern and southern declinations.  Observations 
of the northern stars were carried out with the 9.2-m Hobby-Eberly Telescope (HET) and the High 
Resolution Spectrograph (HRS) at the McDonald Observatory.  Eleven hours of queue observing time 
were allocated for this project by the National Optical Astronomy Observatory (NOAO) by way of the 
Telescope System Instrumentation Program (TSIP).  Seven stars were observed on 13 separate nights, 
with four of the stars being observed on multiple nights.  The HRS fiber-fed echelle spectrograph 
was configured with a standard configuration, using the central echelle and 316g5936 (316 grooves 
mm$^{-1}$ and central wavelength $\lambda = 5936$ {\AA}) cross disperser settings.  The 2\arcsec\ 
fiber was used with no accompanying sky fiber, no image slicer, and no iodine gas cell.  The 
detector is a 4096 $\times$ 4096 two E2V (2048 $\times$ 4096; 15$\micron$ pixels) ccd mosaic 
providing nearly complete spectral coverage from 4660 -- 5920 {\AA} over 27 orders and 6060 -- 
7790 {\AA} over 22 orders, with the inter-ccd spacing accounting for the 140 {\AA} gap.  Two pixel 
binning was used in the cross dispersion direction, while no binning was used in the 
dispersion direction.  To achieve the highest spectral resolution possible, the effective 
slit width was set to 0.25\arcsec\  (projected to 2.1 pixels), providing a nominal resolution 
of $R = 120,000$.  The actual achieved resolution, as measured by small emission 
features in the ThAr comparison spectra, is $R \approx 115,000$.  Total exposure times ranged from
24 -- 165 minutes, resulting in signal-to-noise (S/N) ratios of 600 -- 800.

\begin{figure}
\epsscale{0.9}
\plotone{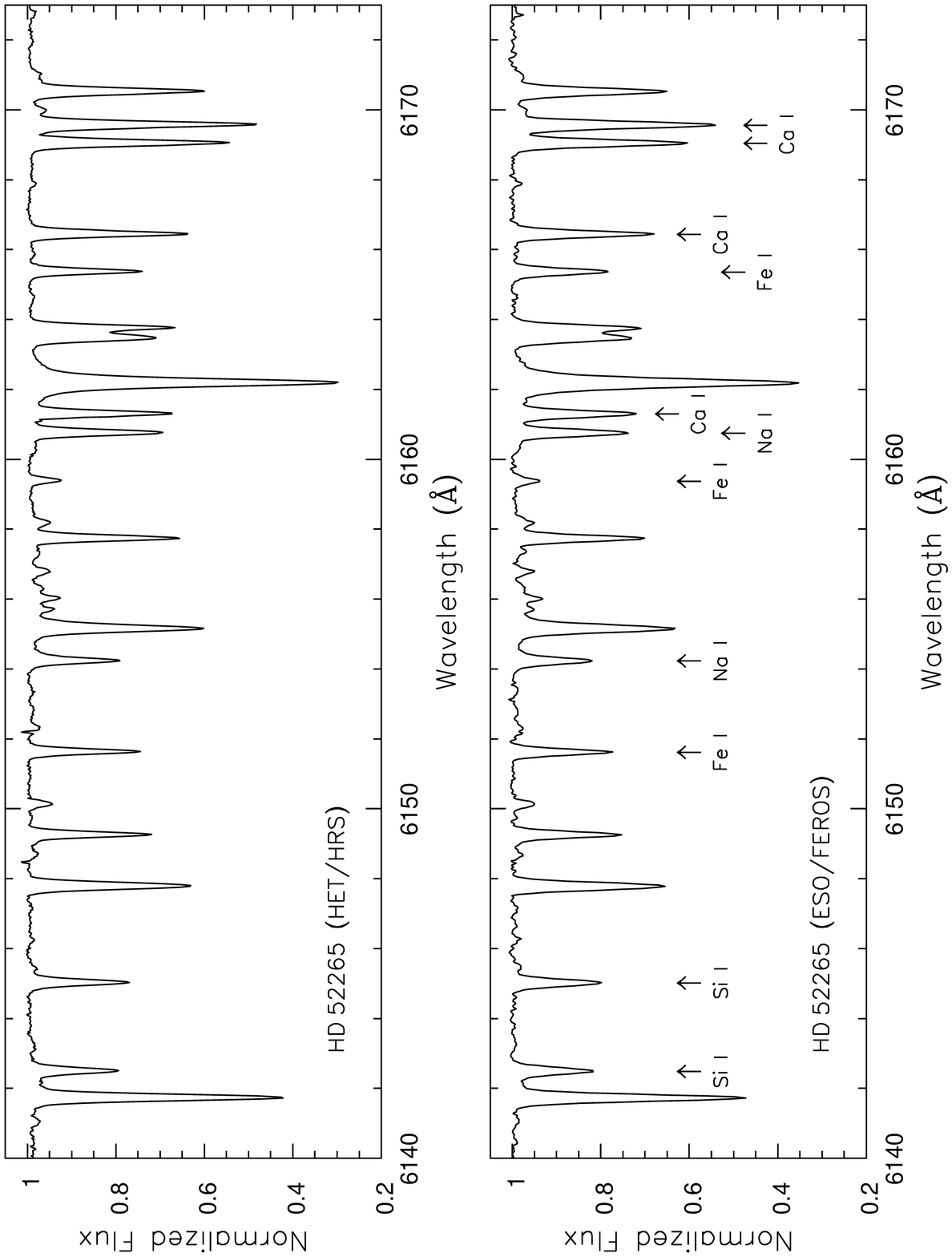}
\caption{\label{fig:spec} Sample spectra of HD\,52265 obtained with HET/HRS (top) and ESO/FEROS 
(bottom).  Lines for which EWs were measured are marked.}
\end{figure}

High-resolution echelle spectra of the southern targets were obtained with 
the 2.2-m MPG/ESO telescope and the Fiber-fed Extended Range Optical Spectrograph (FEROS) at 
the European Southern Observatory (ESO), La Silla under the agreement ESO-Observat{\'o}rio
Nacional/MCT.  These spectra have been used to determine stellar parameters, metallicities, and 
Li abundances of planetary host stars as presented by 
\citet{2010ApJ...720.1290G,2010ApJ...725..721G,2010ApJ...724..154G}, which should be consulted 
for a complete description of the observations and instrumental configuration.  The ESO/FEROS 
spectra have an almost complete spectral coverage from 3560 -- 9200 {\AA} over 39 echelle orders, 
and are characterized by a nominal resolution of $R \sim 48,000$ and signal-to-noise (S/N) ratios 
of 330 -- 400 at 6700 {\AA}.  All of the observations are summarized in an observing log 
presented in Table \ref{tab:obs}, and sample HET/HRS and ESO/FEROS spectra are given in Figure 
\ref{fig:spec}. 

Data reduction was carried out separately for each dataset.  The HET/HRS spectra were reduced
using standard techniques within the IRAF\footnotemark[9] image processing software.  Calibration 
frames (biases, flat fields, ThAr comparison lamps, and telluric standards) were taken on every 
night our objects were observed as part of the observatory's standard calibration plan.  The
reduction process included bias removal, scattered light subtraction, flat fielding, order
extraction, and wavelength calibration. The FEROS Data Reduction System (DRS) was used to reduce 
the ESO/FEROS spectra, with the details provided by \citet{2010ApJ...720.1290G}.

\footnotetext[9]{IRAF is distributed by the National Optical Astronomy Observatory, which is 
operated by the Association of Universities for Research in Astronomy, Inc., under cooperative 
agreement with the National Science Foundation.}

\section{ABUNDANCE ANALYSIS}
\label{s:analysis}
The analysis of our high-quality data included spectroscopically determining stellar parameters
(\teff, $\log g$, and microturbulence [$\xi$]) and deriving the abundances of 18 elements 
spanning condensation temperatures of 40 -- 1659 K for each star.  Abundances have been derived
directly from equivalent width (EW) measurements of spectral lines and by the spectral synthesis
method, depending on the spectral line being considered.  The majority of EWs were measured by 
fitting Gaussian profiles to the lines, whereas some features, generally strong (EW $\geq 90$ 
m{\AA}) lines with broader wings at the continuum, were fit with Voigt profiles.  All EWs were 
measured using the one-dimensional spectrum analysis package SPECTRE \citep{1987BAAS...19.1129F}. 

Abundances from the EW measurements and synthetic fits to the data were derived using an updated
version of the LTE spectral analysis code MOOG \citep{1973ApJ...184..839S}.  Model atmospheres
have been interpolated from the Kurucz ATLAS9 grids\footnotemark[10] constructed assuming the
convective overshoot approximation.  The overshoot models are preferred, because of the finer
grid steps available compared to the more up to date models with no overshoot and new opacity
distribution functions.  Furthermore, no significant differences ($\leq 0.05$ dex) have been 
identified between abundances derived using the overshoot and no overshoot models for 
solar-metallicity open cluster dwarfs \citep[e.g.,][]{2010PASP..122..766S}, so the use of the
overshoot models is not expected here to be an important source of error in the derived 
parameters and abundances.

\footnotetext[10]{See http://kurucz.harvard.edu/grids.html.}

\subsection{Stellar Parameters}
\label{ss:params}
Stellar parameters for each star were derived using standard techniques.  After adopting initial
parameters from the literature \citep{2004A&A...415.1153S}, \teff, $\log g$, $\xi$, and [Fe/H] 
were altered and new [Fe/H] abundances derived until there existed zero correlation between 
[\ion{Fe}{1}/H] and lower excitation potential ($\chi$), and [\ion{Fe}{1}/H] and reduced EW [$\log 
(\mathrm{EW}/\lambda$)], and also the [Fe/H] abundances derived from \ion{Fe}{1} and \ion{Fe}{2} 
lines were equal to within two significant digits.  We note that unique solutions of \teff\ and
$\xi$ are achieved only if there is no ab initio correlation between $\chi$ and EW of the 
\ion{Fe}{1} lines analyzed; no significant correlation exists for our linelist and measured EWs.  
The Fe lines measured were initially chosen from the extensive line list of 
\citet{1990A&AS...82..179T}.  Each case `a' line was then visually inspected in a high-quality 
HET/HRS solar proxy spectrum (daytime sky spectrum; S/N$\sim500$ at $\sim6700$ {\AA}) for potential 
blending, cosmic ray contamination, proximity to order edges, or any other defect they may prevent 
the accurate measurement of a line.  This process resulted in a final Fe line list containing 61 
\ion{Fe}{1} and 11 \ion{Fe}{2} lines.  We note that not all lines were measurable for each star in 
the sample.  Atomic parameters ($\chi$ and transition probabilities [$\log gf$]) were obtained from 
the Vienna Atomic Line Database 
\citep[VALD;][]{1995A&AS..112..525P, 1999A&AS..138..119K, 1999PhScr..T83..162} via email query.  
The [Fe/H] abundances of the target stars were normalized to solar values on a line-by-line 
basis.  The line list with the adopted atomic parameters, and the EW measurements and resulting 
absolute abundances [$\log $ N(Fe)] for each star and the Sun are given in Tables \ref{tab:linesHa} 
and \ref{tab:linesHb} for those observed with HET/HRS and Table \ref{tab:linesF} for those observed 
with ESO/FEROS.

Uncertainties in the stellar parameters are calculated by forcing $1\sigma$ correlations in the
relations described above.  For \teff, the uncertainty is the temperature change required to 
produce a correlation coefficient in [\ion{Fe}{1}/H] vs $\chi$ significant at the $1\sigma$ 
level, and similarly for $\xi$, the correlation between [\ion{Fe}{1}/H] and the reduced EW.  
Determining the uncertainty in $\log g$ requires an iterative process, as thoroughly described 
in \citet{2010AJ....140..293B}.  Briefly, because the difference in the \ion{Fe}{1} and 
\ion{Fe}{2} abundances is sensitive to changes in $\log g$, the uncertainty $\log g$ is related 
to the uncertainty in the Fe abundances.  Accordingly, $\log g$ is altered until the difference in 
the [\ion{Fe}{1}/H] and [\ion{Fe}{2}/H] abundances equals the combined uncertainty in 
[\ion{Fe}{1}/H] and [\ion{Fe}{2}/H], which is the quadratic sum of the uncertainties in each 
individual abundance due to the adopted \teff\ and $\xi$, as well as in the uncertainty in the 
mean ($\sigma_{\mu}$\footnotemark[11]) \ion{Fe}{1} and \ion{Fe}{2} abundances (the derivation of 
the abundance uncertainties is described below).  The method is then iterated, this time 
propagating the initial difference in $\log g$ into the Fe abundance uncertainties.  The final 
uncertainty in $\log g$ is then the difference between the adopted value and the one obtained 
from this second iteration. 

The final parameters and their $1\sigma$ uncertainties are provided in Table \ref{tab:params}. 
Also included in the table are the derived [\ion{Fe}{1}/H] and [\ion{Fe}{2}/H] abundances, along
with the number of lines measured for each and the uncertainty in the mean abundances.

\footnotetext[11]{$\sigma_{\mu} = \sigma / \sqrt{N-1}$, where $\sigma$ is the standard deviation 
and $N$ the number of lines measured.}

\subsection{Abundances}
\label{ss:elements}
Lines for elements other than Fe were identified initially from \citet{1990A&AS...82..179T}. 
Again, each line was inspected visually in our high-quality solar spectrum for blends and other 
defects, and only those that were deemed clean were included in the final line list.  Additional
sources were used for some elements to supplement the initial list: \citet{2005A&A...431..693A}
for \ion{C}{1}; \citet{2007A&A...461..261M} for \ion{Ca}{1}; \citet{2010A&A...516A..46M} for 
\ion{Ti}{1} and \ion{Ti}{2}; \citet{2009ApJ...701.1519R} for \ion{Ti}{2}; 
\citet{2006A&A...449..723G} for \ion{Mn}{1} and \ion{Co}{1}, and \citet{2004A&A...426..619E} for 
\ion{Zn}{1}.  Unless noted below, atomic parameters for all of the lines analyzed are from VALD.  
The final line list, including each line's wavelength, $\chi$, and $\log gf$, and the measured 
EW and derived absolute abundance [$\log N$(m)] for each star are provided in Tables 
\ref{tab:linesHa} and \ref{tab:linesHb} for those observed with HET/HRS and Table \ref{tab:linesF} 
for those observed with ESO/FEROS.  Below we describe the procedures used for those elements that 
required additional attention beyond a direct EW analysis.

\subsubsection{Carbon}
\label{sss:car}
Carbon abundances have been derived from atomic \ion{C}{1} and molecular \swan\ features.  The
\ion{C}{1} lines all arise from high-excitation ($\chi = $7.68 -- 8.65 eV) transitions and thus 
are expected to be susceptible to NLTE effects \citep[e.g.,][]{2005ARA&A..43..481A}.  However, 
the two lines from the lowest energy levels considered here ($\lambda 5052$ and $\lambda5380$) 
have been shown to deviate only slightly from LTE in the Sun and have estimated NLTE corrections
$\leq 0.05$ dex \citep{2005A&A...431..693A}.  \citet{2005PASJ...57...65T}  have investigated NLTE
corrections for these \ion{C}{1} lines in 160 solar-type stars, with $5000 \leq T_{\mathrm{eff}}
\leq 7000$ K, and found the NLTE corrections on par with those found for the Sun, i.e., 
$\leq 0.05$ dex.  The stars in our sample are physically (\teff, $\log g$, [Fe/H]) similar to those 
in the Takeda \& Honda study, and thus comparably small NLTE corrections are expected for 
them.  Consequently, any deviation from LTE should be negated in the solar-normalized [C/H] 
abundances derived from these lines.

\citet{2005ARA&A..43..481A} suggests that \ion{C}{1} lines arising from higher energy levels,
including the remaining three ($\lambda6588$, $\lambda7111$, and $\lambda7113$) in our line list,
should be more sensitive to NLTE effects; however, \citet{2005A&A...431..693A} find corrections 
that are comparable to those for the $\lambda 5052$ and $\lambda5380$ lines for the Sun. All of 
the \ion{C}{1} lines analyzed here give comparable abundances for each star in our sample, with 
typical standard deviations of about 0.04 dex, except for HD\,217107.  For HD\,217107, the two
lower $\chi$ lines have a mean abundance [C/H] $= 0.290 \pm 0.028$ (standard deviation), while the
three higher $\chi$ lines have [C/H] $=0.463 \pm 0.031$ (s.d.).  Measurement error is an unlikely
source of the difference in these abundances given the quality of the data; NLTE effects are a
more likely cause.  HD\,217107 is the most metal-rich star in our sample, and the 
\citet{2005A&A...431..693A} results for the Sun may not be directly applicable to this star.  We
thus adopt the abundance from the two lower $\chi$ lines.  We note that HD\,76700 has a similarly
high metallicity, as well as similar \teff\ and $\log g$, to HD\,217107, and it does not
demonstrate the discrepancy between the lower and higher $\chi$ lines.  However, only one of the
higher $\chi$ lines ($\lambda6588$) was measurable for this star.  The \ion{C}{1} lines analyzed,
the EW measurements, and the absolute abundances are provided in Tables \ref{tab:linesHa},
\ref{tab:linesHb}, and \ref{tab:linesF}.

The \swan\ lines at $\lambda =$ 5086.3 and 5135.6 {\AA} were also analyzed for abundances.  These
features are blends of multiple components of the \swan\ system, so spectral synthesis was used 
for the abundance derivations.  The line list is composed of atomic data from VALD and \swan\ 
molecular data from \citet{1981ApJ...248..228L}; the latter has been modified slightly from the
original in order to fit the features in the Kurucz solar flux atlas \citep{K84} assuming a solar 
abundances of $\log N_{\odot}(\mathrm{C}) = 8.39$ \citep{Asp05}.  A \swan\ dissociation energy of
$D_{0} = 6.297$ eV was assumed.  The syntheses were smoothed to the appropriate resolution using a 
Gaussian broadening function; small unblended lines in the $\lambda5086$ and $\lambda5135$ regions 
were used to determine the full width half maxima (FWHM) of the Gaussian functions.  Best fits of 
the synthesized spectra to the observed spectra were determined by eye.

Solar C abundances were derived by analyzing in the same way the \swan\ features in our solar
spectra, and the \swan-based solar-normalized abundances for each star are in excellent agreement 
with the abundances derived from the high-excitation \ion{C}{1} lines, with differences $\leq 
0.01$ dex for the majority of the stars.  The final adopted [C/H] abundances are the mean values 
of the individual \ion{C}{1}- and \swan-based abundances for each line analyzed.  A comparison of 
the derived C abundances is provided in Table \ref{tab:cando}.

\subsubsection{Nitrogen}
\label{sss:nit}
Nitrogen abundances were determined from spectral synthesis of the $\lambda6703.9$ and 
$\lambda6704.0$ blend and the blend of $\lambda6706.6$ CN features in the $\lambda6707$ 
\ion{Li}{1} region of our spectra.  The Li line list from \citet{K97} was revised and augmented 
with the CN data from \citet{MGM}.  A CN dissociation energy of $D_{0} = 7.65$ eV was assumed and 
the oscillator strengths of the features were adjusted to match the solar flux spectrum 
\citep{K84} with the input solar abundances of $\log N({\rm C}) = 8.39$ and $\log N({\rm N}) = 
7.78$ \citep{Asp05}.   We note that our N abundances are differentially determined: the adopted 
solar abundance is used to calibrate the CN line list, and the resulting stellar N abundances are 
normalized with this same solar value. Concomitantly, these solar-normalized N abundances 
resulting from the weak features we utilize are independent of $\log gf$ value and the adopted 
solar C and N abundances.

Syntheses with varying N abundance were carried out using the mean C abundances described above 
and assuming an input Fe abundance corresponding to the mean value of [Fe/H]; this input Fe 
abundance was converted to an input absolute abundance assuming a solar value of $\log N{\rm 
(Fe)} = 7.52$ \citep[adopted by MOOG; see][]{1991AJ....102.2001S}.  The resulting syntheses were 
smoothed using a rotational broadening function and $v$ sin $i$ values from the literature, as 
well as a Gaussian broadening function to mimic instrumental broadening; the Gaussian FWHM was 
measured from unblended, well-defined emission features in Th-Ar calibration spectra.  We also 
assumed macroturbulent broadening, which was set by forcing the synthetic line depths of the 
$\lambda6703.5, \lambda6704.5, \lambda6705.1$, and $\lambda6710.3$ \ion{Fe}{1} features to match, 
overall, the observed depths after the rotational and instrumental broadening were fixed.

N abundances were determined by minimizing the $\chi^{2}$ values associated with the fit to the 
CN features.  For the majority of our stars, only upper limits on the N abundance could be 
determined.  The final N abundances and uncertainties are given in Tables \ref{tab:absa} and 
\ref{tab:absb}.

\subsubsection{Oxygen}
\label{sss:oxy}
Oxygen abundances have been derived from the forbidden [\ion{O}{1}] line at $\lambda = 6300.3$
{\AA} and the high-excitation \ion{O}{1} triplet at $\lambda = 7771.9, 7774.2,$ and 7775.4 
{\AA}.  Whereas the formation of the $\lambda6300$ [\ion{O}{1}] is well described by LTE 
\citep[e.g.,][]{2003A&A...402..343T}, the \ion{O}{1} triplet is highly sensitive to NLTE effects 
\citep[e.g.,][]{1991A&A...245L...9K,2005ARA&A..43..481A}.  The extent of the effects has been 
shown to be dependent on metallicity, \teff, and $\log g$ 
\citep[e.g.][]{1992A&A...261..255N,2003A&A...402..343T}, with deviations from LTE becoming more
severe for more metal-poor stars, increasing \teff, and decreasing $\log g$.  The physical
parameter space populated by some stars in our sample is such that NLTE effects are expected to be
non-negligible, and for this reason, preference is given to [\ion{O}{1}]-based abundances when
possible. 

Oxygen abundances were derived from the $\lambda6300$ [\ion{O}{1}] line using measured EWs and
the {\sf blends} driver in the MOOG package.  By providing a line list that includes the blending
lines and input abundances for the blending species, the {\sf blends} driver accounts for the
blending lines' contribution to the overall line strength of the feature when calculating the
abundance of the primary element.  In the case of the $\lambda6300$ [\ion{O}{1}] line, the 
blending feature is a \ion{Ni}{1} line consisting of two isotopic components 
\citep{2003ApJ...584L.107J}; here we adopt the weighted $\log gf$ values of the individual 
components as calculated by \citet{2004A&A...415..155B}.  For the [\ion{O}{1}] line, we adopt the 
$\log gf$ value from the careful analysis of \citet{2001ApJ...556L..63A}.  Spectral synthesis 
was also used for some stars to verify continuum placement and the {\sf blends} results.  The 
solar O abundance was derived from the [\ion{O}{1}] line in the same way as the rest of the 
sample.  However, the line in the HET/HRS solar spectrum is unusable due to obliteration by 
atmospheric emission.  Therefore, the [\ion{O}{1}] abundances of the stars observed with HET/HRS 
are normalized using a solar abundance of $\log N_{\odot}({\rm O}) = 8.69$, the abundance derived 
in a previous study \citep{2006AJ....131.1057S} from a high-quality ($R = 60,000$ and S/N 
$\sim 950$) daytime sky spectrum obtained with the Harlan J. Smith 2.7-m telescope.  This 
spectrum is of higher quality than our ESO/FEROS solar spectrum and thus more comparable to our 
HET/HRS spectra.  The measured EWs and absolute abundances of the [\ion{O}{1}] line for the stars 
and the Sun are provided in Tables \ref{tab:linesHa}, \ref{tab:linesHb}, and \ref{tab:linesF}.

The \ion{O}{1} triplet abundances were derived via an EW analysis assuming LTE.  NLTE corrections 
from \citet{2003A&A...402..343T}, which provides an analytical formula to calculate the corrections
for each line of the triplet, were applied to the LTE abundances of each star and the Sun.  The 
\citet{2003A&A...402..343T} formula has the functional form $\Delta = a10^{(b)({\rm EW})}$, were $a$ 
and $b$ are coefficients that are functions of \teff\ and $\log g$.  Coefficients for these
parameters that best match those of our sample stars were chosen.  The resulting NLTE abundances 
are used primarily as a check of the [\ion{O}{1}]-based abundances, but in the cases of HD\,52265 
and HD\,89744, for which [\ion{O}{1}] abundances are not available, the NLTE triplet abundances 
are adopted.  The measured EWs and absolute abundances of the \ion{O}{1} triplet lines are 
provided in Tables \ref{tab:linesHa}, \ref{tab:linesHb}, and \ref{tab:linesF}.

A comparison of the derived O abundances is shown in Table \ref{tab:cando}.  The agreement between 
the [\ion{O}{1}] and NLTE triplet abundances is quite good; the differences are $\leq 0.05$ dex.  
This agreement provides confidence that the NLTE abundances adopted for HD\,52265 and HD\,89744 
are reasonable.

\subsubsection{Odd-$Z$ Elements: Sc, V, Mn, and Co}
\label{sss:hyperfine}
For some odd-$Z$ elements, electron-nucleus interactions can lead to significant hyperfine
structure (hfs) in some transitions.  The splitting of energy levels resulting from the hfs can
produce increased line strengths that, if not properly treated, will lead to overestimated
abundances \citep{2000ApJ...537L..57P}.  Of the elements considered here, Sc, V, Mn, and Co are
susceptible to the hfs, and as such, we have tested the EW-based abundances for these elements by
using spectral synthesis incorporating hfs components to fit one Mn line and two lines each of
Sc, V, and Co.  The measured EWs and the non-hfs absolute abundances of these elements are 
provided in Tables \ref{tab:linesHa}, \ref{tab:linesHb}, and \ref{tab:linesF}, where the lines 
used for the hfs tests are marked.

The hfs components for the four elements are taken from \citet{2006ApJ...640..801J}, and the line 
lists for the regions surrounding each feature were obtained from VALD.  The synthetic spectra
were smoothed using a Gaussian broadening function, and the best fits to the observed spectra were
again determined by eye.  The same analysis was carried out for each solar spectrum, and the
resulting solar abundances were used to normalize the hfs abundances of the stellar sample.  
Results from the hfs syntheses and comparisons to the EW-based abundances indicate that the 
differences between the two abundance determinations are negligible ($\leq 0.04$ dex) for most 
stars.  The two exceptions are the V and Mn abundances of HD\,76700 and HD\,217107, the two most 
metal-rich stars in the sample.  Whereas the majority ($\sim80$\%) EWs for the four elements are 
$\leq 40$ m{\AA} for each star, V and Mn lines have EWs $> 60$ m{\AA} and up to $\sim100$ m{\AA} 
for HD\,76700 and HD\,217107, line strengths that would be expected to have significant hfs 
\citep[e.g.,][]{2000ApJ...537L..57P}.  The final adopted Sc, V, Mn, and Co abundances of all 
stars are those derived from the hfs analysis.

\subsubsection{Abundance Uncertainties}
\label{sss:uncern}
Uncertainties in the derived abundances arise due to errors in the adopted stellar parameters, 
as well as in the spread in abundances derived from individual lines of an element.  The
abundance uncertainties due to the stellar parameters are determined by first calculating the
abundances sensitivities to the adopted parameters.  Sensitivities were calculated for changes of 
$\pm 150$ K in \teff, $\pm 0.25$ dex in $\log g$, and $\pm 0.30 {\rm km s}^{-1}$ in $\xi$.  In
Table \ref{tab:sens} we provide the abundance sensitivities for two stars, HD\,20367 and HD\,76700,
as representative of the sample.  We note that these two stars were observed with HET/HRS and
ESO/FEROS, respectively.  The abundance uncertainty due to each parameter is calculated by then
scaling the sensitivities by the respective parameter uncertainty.  The final total internal 
uncertainties ($\sigma_{\rm tot}$) are the quadratic sum of the individual parameter uncertainties
and the uncertainty in the mean, $\sigma_{\mu}$, for those abundances derived from more than one
line.

For N, three general contributions to the uncertainties in the derived abundances were considered:
fitting uncertainties (which are well-determined given the $\chi^{2}$ approach and assumptions 
about the continuum level uncertainty), the direct effect of parameter errors on the N abundance
itself (as described above), and the effect of uncertainties in the C abundances (which is a 
fixed input in the N analysis) on the derived N abundances.  The final N abundance uncertainties 
are dominated by the direct effect of the $T_{\rm eff}$ uncertainty on the N abundance itself.  
The fitting uncertainties and the effect of uncertainties in $\log g$ on the input mean C 
abundance are also non-negligible contributors to the final total N uncertainties.

\section{RESULTS \& DISCUSSION}
\label{s:RandD}
The solar-normalized abundances and their uncertainties ($\sigma_{\rm tot}$) for the stars observed
with HET/HRS are provided in Table \ref{tab:absa} and those observed with ESO/FEROS in Table 
\ref{tab:absb}.  The uncertainties are all $\leq 0.10$ dex and in most cases are $\leq 0.05$ dex.  
A major factor in the low uncertainties is the collectively small standard deviations in the mean 
abundances-- a testament to the quality of the spectra-- for those elements derived from multiple 
lines.  Also, the sensitivities of the abundances to changes in the stellar parameters are also 
relatively modest for most elements (Table \ref{tab:sens}).

Despite carrying out a homogeneous abundance analysis on the HET/HRS and ESO/FEROS data,
differences in data quality and reduction techniques may lead to disparate abundance derivations.
Results for HD\,52265, the star observed by both telescopes, suggest that this is not the case
here.  The HET/HRS and ESO/FEROS abundances of this star are in excellent agreement, with a mean
difference of $0.03 \pm 0.02$ (s.d.) dex.  This is further support that our abundances 
are good to the $\sim0.05$ dex level. 

Abundances of the stars in our sample have been reported by numerous groups 
\citep[e.g.,][]{1999PASJ...51..505S,2000AJ....119..390G,2001AJ....121..432G,2001PASJ...53.1211T,
2004A&A...415.1153S,2005ChJAA...5..619H,2006MNRAS.370..163B,2006AJ....131.3069L}.  In the
following discussion, we focus on the two papers (and their sources) from which the stars in our 
sample were chosen, namely S01 and E06.

S01 adopted the abundances of 29 stars from \citet{2001AJ....121..432G}
and its preceding companion papers \citep{1998A&A...334..221G,2000AJ....119..390G}.  Our sample
includes four of these stars-- HD\,52265, HD\,89744, HD\,209458, and HD\,217107.  In general, the
abundances from the two analyses are in good agreement, {\it i.e.}, they agree within the combined
uncertainties.  One element that does merit discussion is C, a low-\tc\ element that, along with 
O, heavily influences the slope of the [m/H]-\tc\ relations.  The C abundances of 
\citet{2001AJ....121..432G} are systematically lower than ours by about 0.10 dex, a difference 
that is not statistically significant but one that can dramatically affect the \tc\ slopes.  The 
systematic difference cannot be ascribed to differences in the stellar parameters, nor should the 
difference be due to the adopted $gf$ values since both analyses are done relative to solar 
abundances\footnotemark[12].  Each line list includes five \ion{C}{1} lines, only two of which 
($\lambda$ 5380 and $\lambda6587$) are used by both.  For the two lines in common, the measured 
EWs are in reasonable agreement.  We inspected the three remaining lines ($\lambda 7109$, $\lambda 
7115$, and $\lambda 7117$) used by Gonzalez et al. in our high-quality spectra, and both $\lambda 
7109$ and $\lambda 7115$ appear to be blended with other lines.  The blending is also apparent in 
the Kurucz solar flux atlas \citep{K84}.  We also consulted \citet{1990A&AS...82..179T} and 
\citet{2005A&A...431..693A}, the sources of our \ion{C}{1} line list, and none of the three 
remaining lines appear in those papers, further suggesting the lines may not be suitable for 
precision abundance determinations.  Although the systematic 0.10 dex offset between our C 
abundances and those of \citet{2001AJ....121..432G} cannot be explicitly attributed to the difference 
in the respective line lists, the use of the three blended red \ion{C}{1} lines by Gonzalez et al. 
is a plausible source.

\footnotetext[12]{According to \citet{1997MNRAS.285..403G} and \citet{2000AJ....119..390G}, the 
$gf$ values of the spectral lines used by \citet{2001AJ....121..432G} are determined by an 
inverted analysis of the Sun adopting the solar abundances of \citet[][$\log N(\mathrm{C}) = 
8.56$]{1989GeCoA..53..197A}, and using the Kurucz solar flux atlas \citep{K84} and/or a solar 
reflected spectrum of the asteroid Vesta; it is not clear from these sources which of the solar 
spectra was used for the C lines.}

For the \tc\ analysis of E06, abundances were collected from multiple sources 
\citep{2004A&A...415.1153S,2004A&A...418..703E,2004A&A...426..619E,2005A&A...438..251B,
2006A&A...445..633E,2006A&A...449..723G}.  All ten of the stars in our sample are included in 
these papers, although the same elements were not derived for all of the stars.  The abundances 
used in E06 are in decent agreement with ours, with differences generally less than 0.15 dex and 
within the combined abundance uncertainties.  However, some elements (Al, S, Ca, V, Zn, and Mn) do 
exhibit systematically divergent abundances on the order of $\pm 0.10$ dex for four or more stars.  
Again, differences in the derived stellar parameters cannot account for the systematic abundance 
differences, so the most probable source is other aspects of the abundance analyses, such as 
differences in the line lists, continuum placement, EW measurements, etc.  The abundances of S and Zn 
are of particular interest, because they are both considered volatile elements (\tc\ $< 900$ K) and 
can affect the slope of the [m/H]-\tc\ relations.  The S and Zn abundances reported in 
\citet{2004A&A...426..619E} are systematically lower than ours by about 0.15 and 0.08 dex, 
respectively.  Similarly, in the comparison of their abundances to extant values in the literature, 
their S and Zn abundances are again lower for the majority of the stars 
\citep[][Tables 14 \& 15 therein]{2004A&A...426..619E}.

\subsection{Abundance Trends with \tc-- Volatile and Refractory Elements}
\label{ss:trendsAll} 
\begin{figure}

\plotone{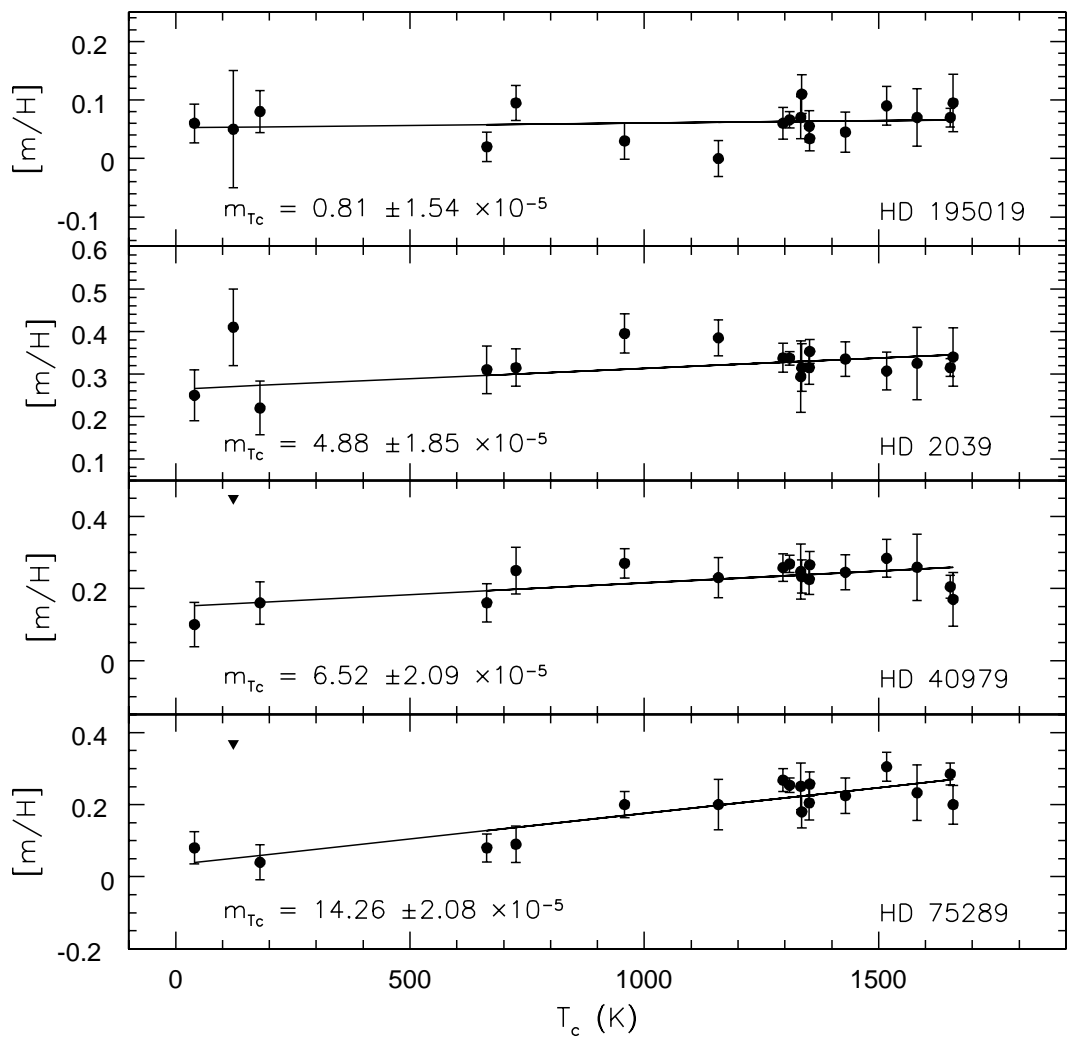}
\caption{\label{fig:tcall} Relative abundances plotted against elemental condensation temperature, 
\tc, for four stars with planets.  The solid line is a linear least-squares fit to the points.  
The slope and uncertainty of the fit are given in the lower left-had corner of each window.  Note 
that the N abundances (\tc\ $= 123$ K) are not included in the linear least-squares fit, as 
described in the text.}
\end{figure}

Similar to previous studies, we quantify the significance of an abundance trend with \tc\ by
the slope of a standard linear least-squares fit.  Fits weighted by the inverse variances of the
solar-normalized abundances have also been made, but to be consistent with the previous studies to 
which our results are compared \citep[S01; E06;][]{2010MNRAS.407..314G}, the unweighted slopes are 
presented and discussed herein.  We note however that the unweighted and weighted slopes for each star 
do not differ significantly, and the conclusions reached in this paper remain unchanged whether the 
unweighted or weighted slopes are considered, indicating that our results are robust.  The fits are 
made to the abundances as a function of the 50\% \tc\ from \citet[][shown here in Table 
\ref{tab:condenTemp}]{2003ApJ...591.1220L} calculated assuming a solar-system composition gas and a 
total pressure of 10$^{-4}$ bar.  The slopes of the fits are given in Table \ref{tab:slopes}, and 
examples are shown in Figure \ref{fig:tcall}.  We note that the derived N abundances are not included 
in the calculation of the slopes because of the larger uncertainty in the N abundances and to maintain 
star-to-star consistency; definitive N measurements were possible for only four of the ten stars.

Positive slopes are found for all ten stars, confirming the results of S01 and E06.  However, for 
all but two stars (HD\,75289 and HD\,76700) our slope measurements are smaller than those of the 
previous studies, in most cases by more than a factor of two.  The differences in the slopes are 
easily understood given the differences in the derived abundances described above.  For example, 
the systematically lower C abundances derived by \citet{2001AJ....121..432G} and used by S01 are 
largely responsible for the more positive slopes of the latter.  Differences in the abundances of 
other elements also contribute to the divergent slopes.

M09 and \citet[][henceforth R09]{2009A&A...508L..17R} have suggested that a precision of 
$\simeq 0.03$ dex in abundance derivations is necessary to detect small differences in trends with 
\tc\ that might distinguish stars with and without planets.  This, they argue, is why previous
studies have not reached strong conclusions about the \tc-dependent abundances of planet host stars.  
This can also explain the differences in the calculated slopes seen here and those of S01 and E06.  
Whereas our abundance uncertainties are $\leq 0.05$ dex, those reported in S01 and E06 are 
typically $\sim 0.10$ dex or higher, resulting in larger uncertainties in the calculated slopes.  
The high quality of our data and the small abundance uncertainties should allow us to make firmer 
conclusions about the [m/H]-\tc\ slopes of our sample stars.

As initially suggested by \citet{1997MNRAS.285..403G}, a positive slope may indicate that the
planetary host star has accreted fractionated rocky material as a consequence of planetary
formation and evolution processes.  Positive slopes also arise from general chemical evolution 
of Galactic disk stars, for instance by the observed trend of decreasing [O/Fe] ratios with
increasing metallicities \citep[e.g.,][]{2007A&A...465..271R}.  The lower O abundances at higher 
metallicities will tend to make more positive the [m/H]-\tc\ relations.  Indeed, S01 (Figure 10
therein) and E06 (Figure 3 therein) demonstrated the effects of chemical evolution on \tc\ slopes by 
comparing slopes of stars with and without known planets as a function of metallicity; both studies 
find a trend of increasing slopes with increasing metallicity, as expected.  The ten stars studied here
were found by S01 and E06 to have slopes that fall above the scatter seen in their respective studies,
and thus were inferred by the authors to have abundance patterns that deviate from those arising from 
general Galactic chemical evolution.  At first sight, confirming the positive slopes for the ten stars 
bolsters the conclusions of S01 and E06 that these stars may have accreted planetary material.  
However, the lower values of the slopes found here for seven stars (HD\,20367, HD\,40979, HD\,52265, 
HD\,89744, HD\,195019, HD\,217107, and HD\,2039) places them in agreement with the Galactic chemical 
evolution trends found by S01 and E06.  While firm conclusions cannot be drawn from a direct 
comparison of our slopes to the Galactic chemical evolution trends defined in S01 and E06 due to 
possible systematic differences arising from the different abundance analyses employed by each study, 
the smaller slopes found here seem to weaken the argument that these stars have accreted substantial 
amounts of planetary material.  For the remaining stars, the slope for HD\,209458 falls near the upper 
envelope of  values for its metallicity, and those for HD\,75289 and HD\,76700 fall appreciably above 
the general Galactic trend.  These three stars, especially the latter two, remain good candidates for 
having accreted fractionated rocky material.

\subsection{Abundance Trends with \tc-- Refractory Elements}
\label{ss:trends900}
R09 showed that the abundance trends of volatile elements in solar twins follow a similar pattern 
as the Sun and that these trends define the general chemical evolution of the Galaxy.  The 
implication is that the Sun and other stars have retained the original volatile composition of the 
proto-stellar nebulae from which they formed.  The abundance trends of the refractory elements 
($T_{\mathrm{c}} > 900$ K), on the other hand, in $\simeq 85 \%$ of solar analogs were found to 
display a strong positive correlation with \tc.  R09 attributed the increasing abundances with 
increasing \tc\ to the composition of refractory elements, which have been shown to be slightly 
depleted relative to volatile elements, in the Sun; this was interpreted as a possible signature 
of terrestrial planet formation in the Solar System (M09).  For the remaining $\simeq 15 \%$ of 
solar analogs, the \tc\ abundance trends of the refractory elements were found to be flat or 
have negative slopes, suggesting their refractory element compositions are more similar to those 
of the Sun and are thus candidates for hosting terrestrial planets.

\begin{figure}
\plotone{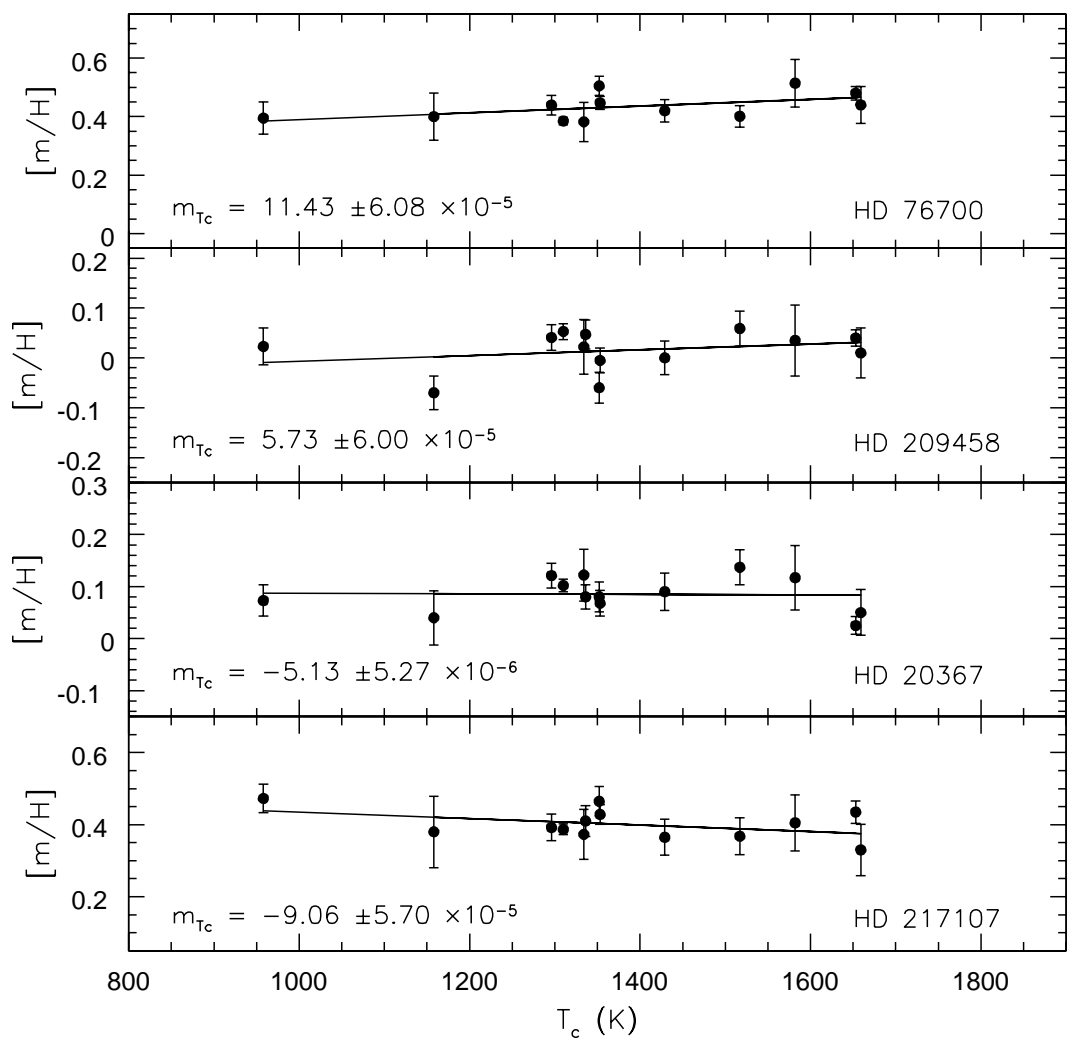}
\caption{\label{fig:tc900} Relative abundances of refractory elements (\tc\ $> 900$ K) plotted 
against elemental condensation temperature for four stars with planets.   The solid line is a 
linear least-squares fit to the points.  The slope and uncertainty of the fit are given in the 
lower left-hand corner of each window.}
\end{figure}

Following R09, we investigate the abundances of refractory elements ($T_{\mathrm{c}} > 900$ K) as 
a function of \tc\ for our sample.  The relations are again quantified by the slope of a standard 
linear least-squares fit to the data.  The slopes of the fits are given in Table \ref{tab:slopes}, 
and examples are shown in Figure \ref{fig:tc900}.  Whereas the [m/H]-\tc\ relations for all elements 
measured have positive slopes for each star, the slopes for the refractory elements seemingly can 
be placed into a group with positive slopes (four stars) and a group with flat or negative slopes 
(six stars).  Of the four stars with positive slopes, one star (HD\,209458) has a slope that is of 
the same order as its uncertainty and thus is also consistent with zero slope.

\noindent
{\bf Positive Slopes:} In the interpretation of R09, stars that display positive [m/H]-\tc\ slopes 
are not terrestrial planet host candidates.  M09 posited that stars with hot Jupiters that do not show 
the solar abundance pattern either accreted their fractionated gas disks while their convection zones 
were still deep and convective mixing erased the planet signature (i.e., enhanced volatiles), or 
interior planets had formed but had been subsequently accreted onto the star, enhancing the 
refractory abundances.  \citet{2010A&A...521A..33R} conclude similarly, suggesting that the 
presence of hot Jupiters prevents the formation of terrestrial planets and consequently the 
appearance of the planet signature, or smaller planets may have already been accreted by the host 
stars.  Future studies will be needed to determine how and if the formation of gas giants affects 
the formation of terrestrial planets; however, the accretion of refractory-rich planet cores may be 
a natural consequence of the constitution of hot Jupiter systems.  \citet{1996Natur.380..606L} showed 
that it is unlikely that gas giant planets can form near (0.05 AU) their host stars, and that hot 
Jupiters formed at larger radii and subsequently migrated to their current locations as a result 
of angular momentum loss via tidal interactions with the surrounding disk (type I migration).  
Migrating gas giants can capture or clear planetary cores along their paths, potentially inducing the 
accretion of at least some of the cores onto the host star \citep{2008ApJ...673..487I}.

\begin{figure}
\plotone{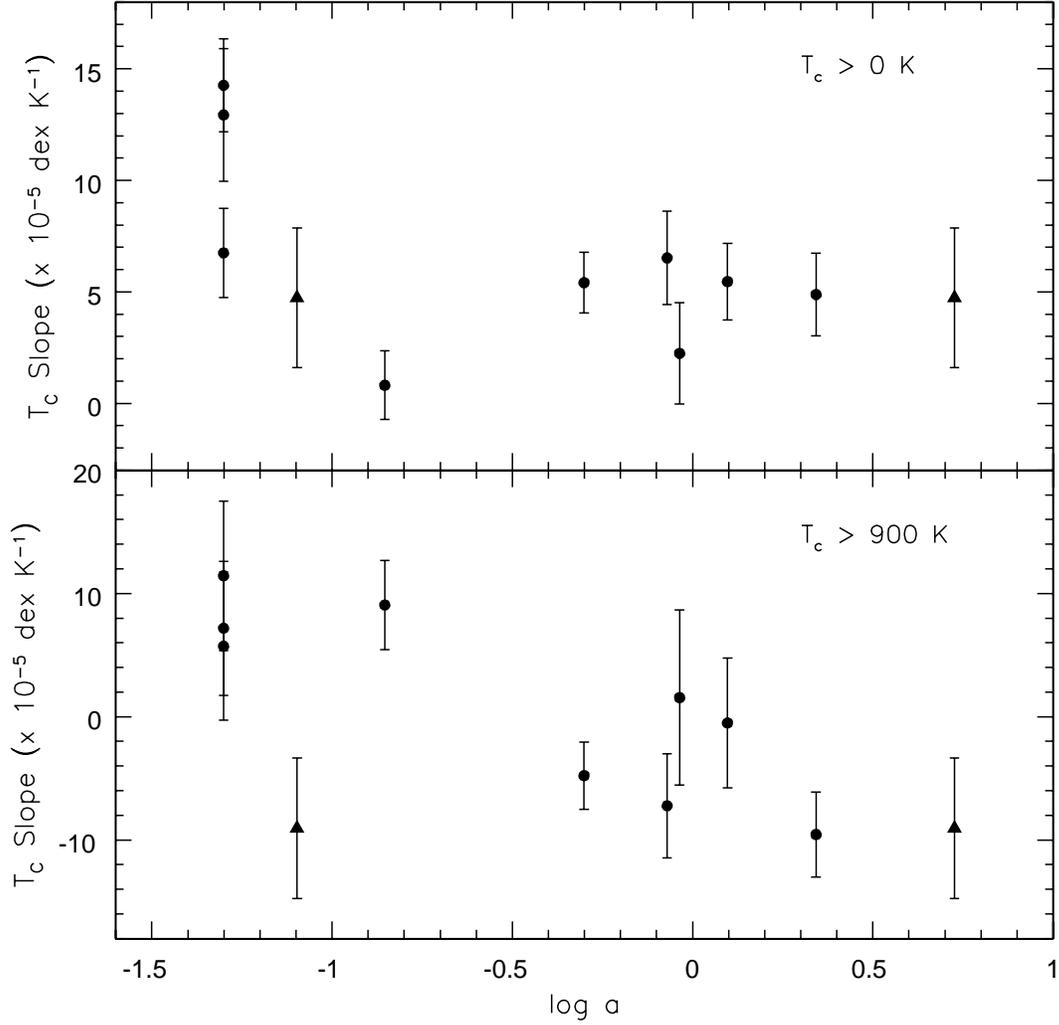}
\caption{\label{fig:tc_axis} \tc\ slope as a function of the log of the semi-major axis (a) of 
the companion planet.  The top panel shows the slopes in the [m/H]-\tc\ relations for all 
elements, and those for the refractory elements only (\tc\ $> 900$ K) are given in the bottom 
panel.  The triangles represent the two planets orbiting HD\,217107.  The error bars represent the 
1-$\sigma$ uncertainties in the slopes given in Table \ref{tab:slopes}.}
\end{figure}

The four stars with positive slopes-- HD\,75289, HD\,76700, HD\,195019, and HD\,209458-- are 
consistent with the accretion scenario.  Properties of the planetary companions of the stars in our
sample are provided in Table \ref{tab:planets}; the planetary data are from the Exoplanet Data
Explorer\footnotemark[13].  As shown in Figure \ref{fig:tc_axis}, the planets with the smallest 
semi-major axes are associated with the four positive slope stars (with exception of HD\,217107 b, 
which is discussed below).  Thus, the stars with the closest-in planets have positive
[m/H]-\tc\ relations for the refractory elements.  Also, HD\,75289, HD\,76700, and HD\,209458, 
when the volatile and refractory elements are considered together, have slope values lying above 
the general Galactic evolution trend (as discussed in Section \ref{ss:trendsAll}).  It seems 
possible that these stars have accreted refractory-rich planet cores.

\footnotetext[13]{Available at http://exoplanets.org}

The magnitudes of the positive slopes found for the four stars are very similar to what would be 
obtained, for example, if $\sim5$ M$_{\oplus}$ of material having the bulk composition of the Earth
\citep[crust, mantle, and core;][]{mcD2001} were mixed into the solar convective envelope ($m\sim0.02$ 
M$_{\odot}$) having a normal solar composition.  Since convective envelope mass is a strong function 
of \teff, stars even slightly hotter than the Sun (say $\sim6000$ K) would require substantially less 
accreted material to create a measurable positive slope.  However, the amount of accreted material 
necessary to produce the derived \tc\ slopes would not increase significantly the overall metallicity 
of the host star, supporting extant evidence that stars hosting giant planets are, on average,
intrinsically more metal-rich than stars not known to host giant planets.

The case of HD\,209458 is particularly interesting.  This star is one of the brightest stars known 
to have a transiting planet, and it has been the focus of intense study. After the discovery of 
HD\,209458 b \citep{2000ApJ...529L..41H,2000ApJ...529L..45C,2000ApJ...532L..55M}, subsequent 
radial velocity \citep{2005ApJ...629L.121L} and transit 
\citep{2007ApJ...658.1328C,2008ApJ...682..586M} searches have not detected additional planets in 
this system.  Also, a search for Trojan-type asteroids found no significance presence of such 
bodies in the system \citep{2010ApJ...716..315M}.  It is not currently possible to know if 
additional planet cores were present when HD\,209458 b formed and migrated to its current orbit, 
but the present lack of planets or other planetary material is intriguing in light of the 
accretion scenario.

\noindent
{\bf Flat or Negative Slopes:} The remaining six stars with flat or negative slopes, in the 
interpretation of R09, are possible hosts of terrestrial planets.  M09 also considered if the 
formation of giant planets could be responsible for the planet signature.  Four solar analogs 
with known close-in giant planets were included in their sample, but all of them were found to have 
abundance patterns that differ from the Sun.  M09 concluded that the presence of close-in giant 
planets is not responsible for the planet signature, per se, and suggested the difference could 
be due to different characteristics of planetary disks giving rise to terrestrial and giant planets.  
However, except for HD\,217107, the five remaining stars in our sample with flat or negative slopes 
are currently known to have only one giant planet not on close-in orbits, with semi-major axes 
ranging from 0.50--2.20 AU (see Table \ref{tab:planets}), so these systems are compatible with the 
alternative explanation of M09.  Also, \citet{2010MNRAS.407..314G} found that stars with giant 
planets have more negative slopes than stars without planets based on a sample of 65 of the former 
and 56 of the latter.  Taken together, these results suggest that the fractionation of volatile and 
refractory elements may be a property of all planetary systems, with the refractory elements being 
locked up in either terrestrial or gas giant planets.

The lone star in our sample that is known to host at least two giant planets, HD\,217107, is also 
consistent with this scenario.  One planet, HD\,217107\,c, is on an extended orbit at 5.33 AU, while 
the second planet, HD\,217107\,b, is on a short orbit at 0.08 AU (Table \ref{tab:planets}).  Despite 
having a close-in giant planet, the negative slope of HD\,217107 implies that significant accretion 
of refractory-rich planet material did not take place in this system as HD\,217107\,b migrated to its 
current location.  This further suggests that terrestrial planets did not form interior to 
HD\,217107\,b, and thus the fractionation of volatile and refractory elements occurs in the formation
of terrestrial and gas giant planets alike.  However, it is also possible that one or more terrestrial 
planets did form interior to HD\,217107\,b but were captured by the larger planet or scattered from 
their original orbits without accreting onto the host star during the planet's migration.  This 
scenario would also conserve the deficiency of refractory elements in the star's photosphere, if in 
fact flat or negative \tc\ slopes result only from the formation of terrestrial planets.

While the flat or negative \tc\ slopes found for six stars in our sample are consistent with the 
planet signature scenario, the abundance trends may be the result of general chemical evolution of the 
Galaxy.  In Figure 5a we plot the \tc\ $> 900$ K slopes as a function of [Fe/H] for the stars in our 
sample.  Included in the figure is the standard linear least-squares fit to the similar slope versus 
[Fe/H] data for stars with and without known giant planets from 
\citet[][Table 1]{2010MNRAS.407..314G}.  The relation is similar to those in R09 and 
\citet{2010A&A...521A..33R}; all of these studies find that the slopes become more negative at higher 
metallicities.  If the relation is indicative of Galactic chemical evolution effects, negative 
slopes in metal-rich stars may not be a signature of planet formation.  As seen in Figure 5a, the 
six stars with flat or negative slopes studied here fall nicely along the fit to the Gonzalez et al. 
data, despite possible systematic differences in the \tc\ slopes between the two studies, and when the 
slopes are corrected for chemical evolution, the effect is clearer (Figure 5b).  Tellingly, three of 
the four stars with close-in planets (HD\,75289, HD\,76700, and HD\,195019) have slopes that lie above 
the Galactic trend by more than $2\sigma$, providing additional evidence that these stars have 
accreted refractory-rich planetary material.  The slope for HD\,209458, the fourth star with a 
close-in planet, also lies above the trend but at a low confidence level ($\sim 1\sigma$).

\section{SUMMARY}
Stellar parameters and abundances of 18 elements have been homogeneously derived for 10 stars 
known to host Jovian-type giant planets.  The LTE analysis is based on high-quality echelle 
spectroscopy obtained with the 9.2-m Hobby Eberly and 2.2-m MPG/ESO telescopes.  Stellar 
parameters were determined spectroscopically using the standard iterative technique.  Abundances 
were derived from measured equivalent widths or synthesis of spectral lines, and have internal 
uncertainties that are typically $\leq 0.05$ dex.  Special attention was given to the derivation 
of the important volatile elements C, N, and O, as well as the odd-$Z$ elements Sc, V, Mn, and Co.  
Carbon abundances were derived from high-excitation \ion{C}{1} and molecular \swan\ lines, and the 
results from both features are in excellent agreement, with uncertainties in the mean abundances 
$\leq 0.03$ dex.  Adopting the derived C abundances, N abundances were determined by analysis of 
three CN features in the $\lambda 6707$ \ion{Li}{1} region.  Definitive measurements were possible 
for only four stars; upper limits are reported for the remaining six.  Oxygen abundances have been 
derived from the $\lambda 6300$ [\ion{O}{1}] forbidden line and the high-excitation \ion{O}{1} 
triplet with NLTE corrections from \citet{2003A&A...402..343T}.  Differences in the abundances from 
the two features are $\leq 0.05$ dex.  Account for hyperfine structure was taken in the derivation 
of Sc, V, Mn, and Co abundances.  In most cases, the effect is less than 0.04 dex on the derived 
abundances; however, for the two most metal-rich stars in the sample, the difference is as high as 
0.36 dex.

We have examined the abundances derived from our fine analysis as a function of condensation 
temperature of the elements to look for trends that may be related to the planet formation process.  
The precision of our abundances ($\leq 0.05$ dex) is of the order necessary to detect the potentially 
small abundance differences that may distinguish stars with and without planets.  When considering 
the volatile and refractory elements together, we find positive slopes in the [m/H]-\tc\ relations
for all ten stars, in agreement with \citet{2001AJ....121.3207S} and \citet{2006A&A...449..809E}. 
The slopes derived here are in general smaller (less positive) than those of S01 and E06 due 
primarily to systematic differences in the derived abundances.  For seven stars, the [m/H]-\tc\ 
slopes fall along the trend of slope versus metallicity that defines the general chemical 
evolution of the Galaxy and thus do not appear to be indicative of planet formation around these 
stars.  The remaining three stars- HD\,75289, HD\,76700, and HD\,209458- have slopes lying above 
the Galactic evolution trend and are candidates for having accreted fractionated rocky material 
during the formation and/or evolution of their planetary systems.

It has been argued that volatile elements are more sensitive to Galactic chemical evolution 
effects than refractory elements and that trends with \tc\ of the latter are more robust 
when looking for a planet signature among stellar abundances \citep{2010A&A...521A..33R}.  The 
slopes of the [m/H]-\tc\ relations for the refractory elements of our sample are dichotomized into 
groups with positive (four stars), and flat or negative (six stars) values.  Positive slopes are a 
possible indication that there was no fractionation of volatile and refractory elements in the 
protoplanetary disks of the stars and thus terrestrial planet formation was suppressed.  
Alternatively, terrestrial planets or planet cores could have formed but were subsequently accreted 
onto the star due to dynamical processes in the disk, causing an enhancement in the photospheric 
abundances of the refractory elements.  The four stars in our sample with positive \tc\ slopes have 
very close-in ($\leq 0.14$ AU) giant planets, which are thought to have migrated to their current 
locations after forming at larger radii.  Three of these stars also have volatile $+$ refractory \tc\ 
slopes lying above the general Galactic evolution trend, and all four lay above the Galactic trend 
for \tc\ $> 900$ K.  These data strengthen the evidence that these four stars have undergone 
accretion of refractory-rich planet material.

Flat or negative \tc\ slopes for the refractory elements have been interpreted as a possible signature 
of terrestrial planet formation \citep{2009ApJ...704L..66M,2009A&A...508L..17R}.  Six stars in our 
sample with flat or negative \tc\ slopes-- HD\,2039, HD\,20367, HD\,40979, HD\,52265, HD\,89744, and 
HD\,217107-- are candidates for hosting terrestrial planets.  However, the planet signature may not 
be limited to the formation of terrestrial planets but may result from the formation of gas giants, 
as well; this is evident by our sample of giant planet hosts.  Furthermore, HD\,217107, is the only 
star in our sample with two known planets; it has a 2.6 $M_{\mathrm{J}}$ planet orbiting at 5.33 AU 
and a 1.4 $M_{\mathrm{J}}$ planet orbiting at 0.08 AU.  The negative \tc\ slope for this star 
suggests that fractionation of the refractory elements did occur and that significant accretion of 
refractory-rich planet material has not taken place despite having a Jovian-type giant planet on a 
close-in orbit.  It seems then that the fractionation of volatile and refractory elements may be 
a process inherent to the formation of terrestrial and gas giant planets alike.  However, 
interpretation of abundance trends may be complicated by Galactic chemical evolution effects.  Larger 
samples of stars with and without known planets subject to a homogeneous abundance analysis based on 
high-quality spectroscopy are needed to determine definitively if the chemical abundance distributions 
of stars with known planets differ from the general stellar population.

\acknowledgements
S.C.S. acknowledges support provided by the NOAO Leo Goldberg Fellowship; NOAO 
is operated by the Association of Universities for Research Astronomy, Inc. (AURA), 
under a cooperative agreement with the National Science Foundation (NSF).  D.F. was 
supported by the NOAO/KPNO Research Experiences for Undergraduates Program which is 
funded by the NSF Research Experiences for Undergraduates (REU) Program and the 
Department of Defense ASSURE program through Scientific Program Order No. 13 
(AST-0754223) of the Cooperative Agreement No. AST-0132798 between AURA and the NSF.  
J.R.K. gratefully acknowledges support for this work by grants AST 00-86576 and AST 
02-39518 to J.R.K. from the National Science Foundation and by a generous grant 
from the Charles Curry Foundation to Clemson University.  L.G. acknowledges the 
financial support from CNPq.  The Hobby-Eberly Telescope (HET) is a joint project 
of the University of Texas at Austin, the Pennsylvania State University,Standford 
University, Ludwig-Maximilians-Universit{\"a}t M{\"u}nchen, and 
Georg-August-Universit{\"a}t G{\"o}ttingen.  The HET is named in honor of its principal
benefactors, William P. Hobby and Robert E. Eberly.  We thank the anonymous referee for helpful
comments that have led to an improved paper.

{\it Facilities:} \facility{HET(HRS)} \facility{ESO:2.2m(FEROS)}

\newpage

\begin{figure}
\plotone{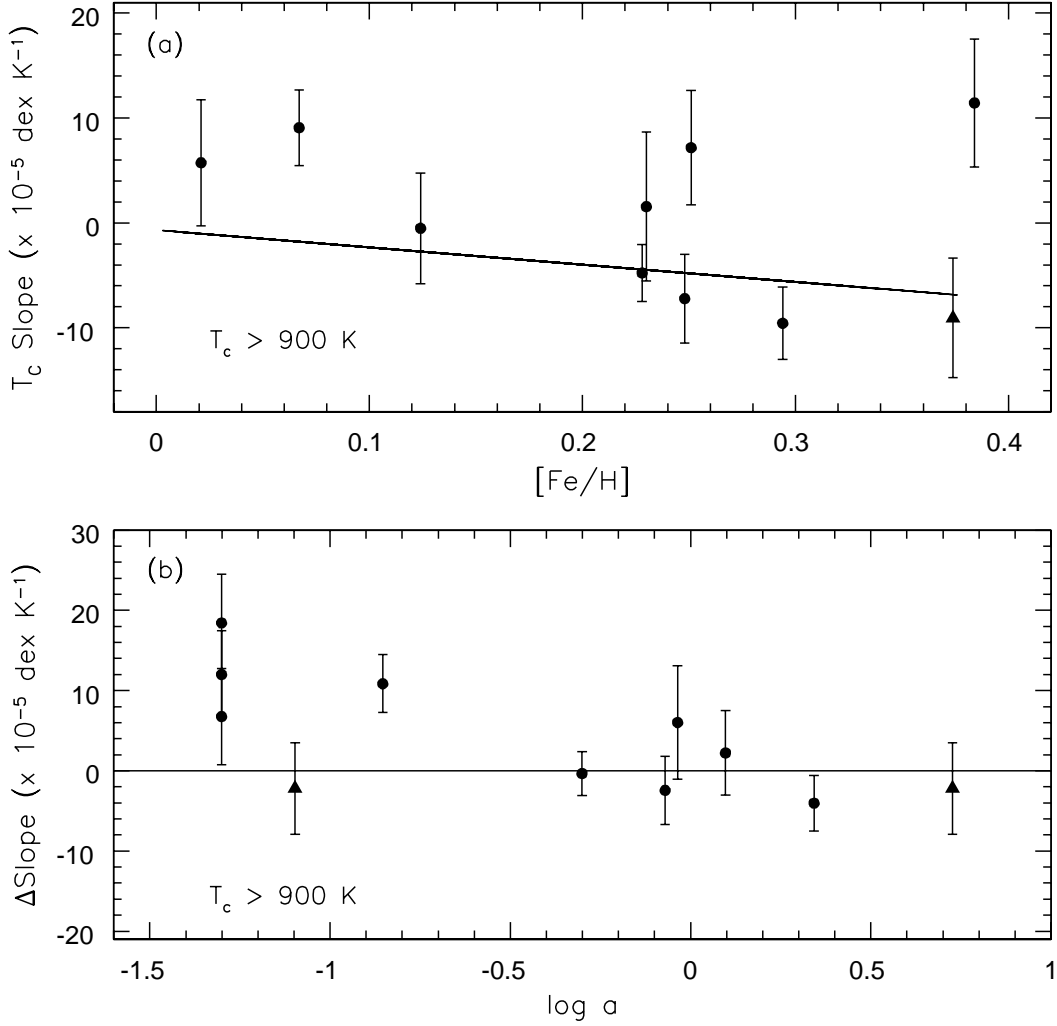}
\caption{\label{fig:tc_feh} (a) \tc\ slope for the refractory elements (\tc\ $> 900$ K) as a function 
of [Fe/H].  HD\,217107, the only star in our sample with two known planets, is given as the triangle.  
The solid line is the linear least-squares fit to the slope-[Fe/H] data for stars with and without 
known planets from \citet{2010MNRAS.407..314G} and defines the Galactic chemical evolution trend. (b)
\tc\ slope for the refractory elements corrected for Galactic chemical evolution versus the log of the 
semi-major axis of the companion planet.  The corrected slopes are the difference between the measured
slope and the [Fe/H]-dependent fitted value for each star from the Galactic chemical evolution trend 
shown in panel (a).  The two known planets of HD\,217107 are again given as triangles.}
\end{figure}

\input{tab1.tex}
\input{tab2a.tex}

\input{tab2b.tex}

\input{tab3.tex}
\input{tab4.tex}

\input{tab5.tex}

\input{tab6a.tex}

\input{tab6b.tex}

\input{tab7.tex}
\input{tab8.tex}
\input{tab9.tex}

\input{tab10.tex}

\end{document}

%% file: tab1.tex
\begin{deluxetable}{lcccccccr}
\tablecolumns{9}
\tablewidth{0pt}
\tablenum{1}
\tablecaption{Observing Log\label{tab:obs}}
\tablehead{
     \colhead{}&
     \colhead{}&
     \colhead{}&
     \colhead{}&
     \colhead{Date}&
     \colhead{}&
     \colhead{}&
     \colhead{}&
     \colhead{$T_{\mathrm{exp}}$}\\
     \colhead{Star}&
     \colhead{}&
     \colhead{$V$}&
     \colhead{Telescope}&
     \colhead{(UT)}&
     \colhead{}&
     \colhead{$N$}&
     \colhead{}&
     \colhead{(s)}
     }
\startdata
HD\,2039   && 9.00 & ESO & 2007 Aug 28 && 2 && 1500\\
HD\,20367  && 6.40 & HET & 2007 Mar 10 && 1 && 1560\\
           &&      &	 & 2007 Sep 28 && 1 && 1342\\
	   &&      &     & 2007 Oct 03 && 2 && 1560\\
HD\,40979  && 6.73 & HET & 2007 Feb 27 && 2 && 1080\\
           &&      &     & 2007 Feb 28 && 2 && 1080\\
HD\,52265  && 6.30 & HET & 2007 Feb 27 && 2 && 1440\\
	   &&      & ESO & 2007 Apr 08 && 2 &&  100\\
HD\,75289  && 6.36 & ESO & 2007 Apr 07 && 2 &&  100\\
HD\,76700  && 8.13 & ESO & 2007 Apr 06 && 2 &&  600\\
HD\,89744  && 5.74 & HET & 2007 Mar 05 && 1 && 1740\\
HD\,195019 && 6.91 & HET & 2007 May 10 && 2 && 1200\\
           &&      &     & 2007 May 14 && 2 && 1200\\
HD\,209458 && 7.65 & HET & 2007 Jun 08 && 2 && 1240\\
           &&      &     & 2007 Jun 21 && 2 && 1240\\
	   &&      &     & 2007 Jul 16 && 2 && 1240\\
	   &&      &     & 2007 Aug 16 && 2 && 1240\\
HD\,217107 && 6.18 & HET & 2007 Aug 10 && 2 && 1260\\
\enddata

\end{deluxetable}

%% file: tab2a.tex
\begin{deluxetable}{lcccccccrccrccrccrc}
\tablecolumns{19}
\tablewidth{0pt}
\tabletypesize{\small}
\rotate
\tablenum{2a}
\tablecaption{Lines Measured, Equivalent Widths, and Abundances- HET/HRS (Sun, HD\,20367, HD\,40979, \& HD\,52265) \label{tab:linesHa}}
\tablehead{
     \colhead{}&
     \colhead{}&
     \colhead{$\lambda$}&
     \colhead{}&
     \colhead{$\chi$}&
     \colhead{}&
     \colhead{}&
     \colhead{}&
     \colhead{}&
     \colhead{}&
     \colhead{}&
     \multicolumn{2}{c}{HD \, 20367}&
     \colhead{}&
     \multicolumn{2}{c}{HD \, 40979}&
     \colhead{}&
     \multicolumn{2}{c}{HD \, 52265}\\
     \colhead{Ion}&
     \colhead{}&
     \colhead{(\AA)}&
     \colhead{}&
     \colhead{(eV)}&
     \colhead{}&
     \colhead{$\log \mathrm{gf}$}&
     \colhead{}&
     \colhead{EW$_{\odot}$}&
     \colhead{$\log N_{\odot}$}&
     \colhead{}&
     \colhead{EW}&
     \colhead{$\log N$}&
     \colhead{}&
     \colhead{EW}&
     \colhead{$\log N$}&
     \colhead{}&
     \colhead{EW}&
     \colhead{$\log N$}
     }
\startdata
\ion{C}{1}   && 5052.17 && 7.68 && -1.304 && 33.1 & 8.46 && 42.1 & 8.44 && 55.0 & 8.57 && 55.8 & 8.60 \\
             && 5380.34 && 7.68 && -1.615 && 21.9 & 8.53 && 27.3 & 8.47 && 39.4 & 8.64 && 40.2 & 8.68 \\
             && 6587.61 && 8.54 && -1.021 && 13.9 & 8.43 && 21.3 & 8.45 && 30.2 & 8.58 && 29.8 & 8.60 \\
             && 7111.47 && 8.64 && -1.074 && 10.1 & 8.41 && 17.7 & 8.49 && 21.7 & 8.52 && 22.1 & 8.56 \\
             && 7113.18 && 8.65 && -0.762 && 22.5 & 8.56 && 27.5 & 8.44 && 39.1 & 8.59 && 40.1 & 8.64 \\
\ion{O}{1}   && 6300.30 && 0.00 && -9.717 &&  5.5\tablenotemark{a} & 8.69\tablenotemark{a} &&  4.3 & 8.71 &&  6.0 & 8.85 &&\nodata&\nodata \\
             && 7771.94 && 9.15 &&  0.369 && 66.8 & 8.80 &&101.7 & 8.94 &&119.2 & 9.08 &&108.8 & 9.00 \\
             && 7774.17 && 9.15 &&  0.223 && 58.8 & 8.81 && 88.8 & 8.93 &&106.3 & 9.07 && 97.4 & 9.00 \\
\enddata

\tablenotetext{a}{Taken from Schuler et al. (2006).}

\tablenotetext{\dagger}{Line used for hfs tests.}

\end{deluxetable}


%% file: tab2b.tex
\begin{deluxetable}{lcccccccrccrccrccrc}
\tablecolumns{19}
\tablewidth{0pt}
\tabletypesize{\small}
\rotate
\tablenum{2b}
\tablecaption{Lines Measured, Equivalent Widths, and Abundances- HET/HRS (HD\,89744, HD\,105019, HD\,209458, \& HD\,217107)\label{tab:linesHb}}
\tablehead{
     \colhead{}&
     \colhead{}&
     \colhead{$\lambda$}&
     \colhead{}&
     \colhead{$\chi$}&
     \colhead{}&
     \colhead{}&
     \colhead{}&
     \multicolumn{2}{c}{HD \, 89744}&
     \colhead{}&
     \multicolumn{2}{c}{HD \, 195019}&
     \colhead{}&
     \multicolumn{2}{c}{HD \, 209458}&
     \colhead{}&
     \multicolumn{2}{c}{HD \, 217107}\\
     \colhead{Ion}&
     \colhead{}&
     \colhead{(\AA)}&
     \colhead{}&
     \colhead{(eV)}&
     \colhead{}&
     \colhead{$\log \mathrm{gf}$}&
     \colhead{}&
     \colhead{EW}&
     \colhead{$\log N$}&
     \colhead{}&
     \colhead{EW}&
     \colhead{$\log N$}&
     \colhead{}&
     \colhead{EW}&
     \colhead{$\log N$}&
     \colhead{}&
     \colhead{EW}&
     \colhead{$\log N$}
     }
\startdata
\ion{C}{1}   && 5052.17 && 7.68 && -1.304 && 74.5 & 8.70 && 43.8 & 8.57 && 40.8 & 8.40 && 46.5 & 8.77\\
             && 5380.34 && 7.68 && -1.615 && 51.9 & 8.69 && 26.6 & 8.55 && 25.0 & 8.41 && 31.4 & 8.80\\
             && 6587.61 && 8.54 && -1.021 && 40.6 & 8.64 && 19.7 & 8.54 &&\nodata&\nodata&& 27.7 & 8.90\\
             && 7111.47 && 8.64 && -1.074 && 31.7 & 8.61 && 13.5 & 8.47 && 13.3 & 8.34 && 19.9 & 8.84\\
             && 7113.18 && 8.65 && -0.762 && 50.0 & 8.63 && 29.6 & 8.65 && 26.7 & 8.43 && 40.9 & 9.05\\
\ion{O}{1}   && 6300.30 && 0.00 && -9.717 &&\nodata&\nodata&&  7.3 & 8.77 &&  4.5 & 8.65 && 10.7 & 8.92\\
             && 7771.94 && 9.15 &&  0.369 &&140.6 & 9.23 && 79.3 & 8.91 && 94.0 & 8.85 && 73.7 & 9.05\\
             && 7774.17 && 9.15 &&  0.223 &&127.4 & 9.21 && 71.0 & 8.93 && 83.7 & 8.86 && 65.8 & 9.06\\
\enddata

\tablenotetext{\dagger}{Line used for hfs tests.}

\end{deluxetable}


%% file: tab3.tex
\begin{deluxetable}{lcccccccrccrccrccrccrc}
\tablecolumns{22}
\tablewidth{0pt}
\tabletypesize{\footnotesize}
\rotate
\tablenum{3}
\tablecaption{Lines Measured, Equivalent Widths, and Abundances- ESO/FEROS\label{tab:linesF}}
\tablehead{
     \colhead{}&
     \colhead{}&
     \colhead{$\lambda$}&
     \colhead{}&
     \colhead{$\chi$}&
     \colhead{}&
     \colhead{}&
     \colhead{}&
     \colhead{}&
     \colhead{}&
     \colhead{}&
     \multicolumn{2}{c}{HD \, 2039}&
     \colhead{}&
     \multicolumn{2}{c}{HD \, 52265}&
     \colhead{}&
     \multicolumn{2}{c}{HD \, 75289}&
     \colhead{}&
     \multicolumn{2}{c}{HD \, 76700}\\
     \colhead{Ion}&
     \colhead{}&
     \colhead{(\AA)}&
     \colhead{}&
     \colhead{(eV)}&
     \colhead{}&
     \colhead{$\log \mathrm{gf}$}&
     \colhead{}&
     \colhead{EW$_{\odot}$}&
     \colhead{$\log N_{\odot}$}&
     \colhead{}&
     \colhead{EW}&
     \colhead{$\log N$}&
     \colhead{}&
     \colhead{EW}&
     \colhead{$\log N$}&
     \colhead{}&
     \colhead{EW}&
     \colhead{$\log N$}&
     \colhead{}&
     \colhead{EW}&
     \colhead{$\log N$}
     }
\startdata
\ion{C}{1}   && 5052.17 && 7.68 && -1.304 && 32.8 & 8.45 && 53.8 & 8.70 && 57.9 & 8.62 && 55.4 & 8.57 && 50.0 & 8.76\\ 
             && 5380.34 && 7.68 && -1.615 && 20.8 & 8.49 && 35.5 & 8.71 && 40.5 & 8.67 && 36.5 & 8.58 && 31.8 & 8.74\\ 
             && 6587.61 && 8.54 && -1.021 && 15.3 & 8.48 && 30.0 & 8.74 &&\nodata&\nodata&& 27.8 & 8.53 && 22.8 & 8.69\\ 
             && 7113.18 && 8.65 && -0.762 && 21.6 & 8.53 &&\nodata&\nodata&&\nodata&\nodata&& 39.2 & 8.61 &&\nodata&\nodata\\ 
\ion{O}{1}   && 6300.30 && 0.00 && -9.717 &&  5.3 & 8.66 &&  7.8 & 8.88 &&\nodata&\nodata&&  5.8 & 8.70 && 10.4 & 8.90\\
             && 7771.94 && 9.15 &&  0.369 && 71.3 & 8.87 && 96.5 & 9.05 &&109.7 & 8.96 &&108.1 & 8.98 && 77.8 & 9.01\\ 
             && 7774.17 && 9.15 &&  0.223 && 60.9 & 8.85 && 88.0 & 9.08 && 97.2 & 8.95 && 94.4 & 8.95 && 68.7 & 9.01\\ 
             && 7775.39 && 9.15 &&  0.001 && 44.6 & 8.78 && 70.4 & 9.05 && 81.5 & 8.96 && 73.5 & 8.87 && 56.8 & 9.02\\
\enddata

\tablenotetext{\dagger}{Line used for hfs tests.}

\end{deluxetable}

%% file: tab4.tex
\begin{deluxetable}{lcccccccccccrcccrc}
\tablecolumns{18}
\rotate
\tablewidth{0pt}
\tablenum{4}
\tablecaption{Stellar Parameters\label{tab:params}}
\tablehead{
     \colhead{}&
     \colhead{}&
     \colhead{$T_{\mathrm{eff}}$}&
     \colhead{$\sigma$}&
     \colhead{}&
     \colhead{}&
     \colhead{}&
     \colhead{}&
     \colhead{$\xi$}&
     \colhead{$\sigma$}&
     \colhead{}&
     \colhead{}&
     \colhead{}&
     \colhead{}&
     \colhead{}&
     \colhead{}&
     \colhead{}&
     \colhead{}\\
     \colhead{Star}&
     \colhead{}&
     \colhead{(K)}&
     \colhead{(K)}&
     \colhead{}&
     \colhead{$\log g$}&
     \colhead{$\sigma$}&
     \colhead{}&
     \colhead{(km s$^{-1}$)}&
     \colhead{(km s$^{-1}$)}&
     \colhead{}&
     \colhead{[\ion{Fe}{1}/H]}&
     \colhead{$N$}&
     \colhead{$\sigma_{\mu}$}&
     \colhead{}&
     \colhead{[\ion{Fe}{2}/H]}&
     \colhead{$N$}&
     \colhead{$\sigma_{\mu}$}
     }
\startdata
HD\,20367  && 6128 & 33 && 4.52 & 0.09 && 1.78 & 0.11 && 0.124 & 61 & 0.005 && 0.119 & 11 & 0.011\\
HD\,40979  && 6205 & 50 && 4.40 & 0.16 && 1.75 & 0.12 && 0.248 & 57 & 0.007 && 0.246 & 10 & 0.018\\
HD\,52265H && 6145 & 33 && 4.35 & 0.14 && 1.80 & 0.09 && 0.228 & 60 & 0.005 && 0.226 & 11 & 0.013\\
HD\,89744  && 6196 & 43 && 3.89 & 0.15 && 1.81 & 0.11 && 0.230 & 49 & 0.007 && 0.231 &  9 & 0.013\\
HD\,195019 && 5787 & 31 && 4.16 & 0.06 && 1.59 & 0.07 && 0.067 & 61 & 0.005 && 0.073 & 11 & 0.006\\
HD\,209458 && 6075 & 33 && 4.37 & 0.11 && 1.86 & 0.14 && 0.021 & 57 & 0.005 && 0.023 & 11 & 0.009\\
HD\,217107 && 5666 & 40 && 4.26 & 0.14 && 1.35 & 0.07 && 0.374 & 50 & 0.007 && 0.372 & 10 & 0.018\\
           &&	   &	&&	&      &&      &      &&       &    &       &&	     &    &      \\
\enddata

\end{deluxetable}

%% file: tab5.tex
\begin{deluxetable}{lcccccccccccrcccccccr}
\tablecolumns{21}
\rotate
\tabletypesize{\small}
\tablewidth{0pt}
\tablenum{5}
\tablecaption{Carbon \& Oxygen Abundances\label{tab:cando}}
\tablehead{
     \colhead{}&
     \colhead{}&
     \colhead{}&
     \colhead{}&
     \colhead{}&
     \colhead{}&
     \colhead{}&
     \colhead{}&
     \colhead{}&
     \colhead{}&
     \colhead{}&
     \colhead{}&
     \colhead{[\ion{O}{1}]}&
     \colhead{}&
     \multicolumn{5}{c}{\ion{O}{1} Triplet}&
     \colhead{}&
     \colhead{}\\
     \cline{13-13} \cline{15-19}\\
     \colhead{Star}&
     \colhead{}&
     \colhead{[\ion{C}{1}/H]}&
     \colhead{$\sigma$}&
     \colhead{}&
     \colhead{[\swan/H]}&
     \colhead{$\sigma$}&
     \colhead{}&
     \colhead{$<[\mathrm{C/H}]>$\tablenotemark{a}}&
     \colhead{$N$}&
     \colhead{$\sigma_{\mu}$}&
     \colhead{}&
     \colhead{[O/H]}&
     \colhead{}&
     \colhead{[O/H]$_{{\rm LTE}}$}&
     \colhead{$\sigma$}&
     \colhead{}&
     \colhead{[O/H]$_{{\rm NLTE}}$}&
     \colhead{$\sigma$}&
     \colhead{}&
     \colhead{[O/H]\tablenotemark{a}}
     }
\startdata
HD\,20367  && -0.02 & 0.08 && +0.01 & 0.07 && -0.01 & 7 & 0.03 && +0.02 && +0.11 & 0.02 && +0.07 & 0.02 && +0.02\\
HD\,40979  && +0.10 & 0.04 && +0.10 & 0.06 && +0.10 & 7 & 0.02 && +0.16 && +0.25 & 0.03 && +0.18 & 0.02 && +0.16\\
HD\,52265H && +0.14 & 0.03 && +0.14 & 0.01 && +0.14 & 7 & 0.01 &&\nodata&& +0.18 & 0.01 && +0.13 & 0.01 && +0.13\\
HD\,89744  && +0.18 & 0.07 &&\nodata&\nodata&&+0.18 & 5 & 0.03 &&\nodata&& +0.40 & 0.02 && +0.22 & 0.01 && +0.22\\
HD\,195019 && +0.08 & 0.04 && +0.02 & 0.01 && +0.06 & 7 & 0.02 && +0.08 && +0.11 & 0.01 && +0.06 & 0.01 && +0.08\\
HD\,209458 && -0.10 & 0.04 && -0.10 & 0.05 && -0.10 & 6 & 0.02 && -0.04 && +0.05 & 0.01 && +0.01 & 0.02 && -0.04\\
HD\,217107 && +0.29 & 0.03 && +0.36 & 0.01 && +0.32 & 7 & 0.02 && +0.23 && +0.24 & 0.01 && +0.21 & 0.01 && +0.23\\
           &&       &      &&       &      &&       &   &      &&	&&	 &	&&	 &	&&	\\
\enddata

\tablenotetext{a}{Final adopted abundances.}

\end{deluxetable}

%% file: tab6a.tex
\begin{deluxetable}{lclllllllllllll}
\tablecolumns{15}
\tablewidth{0pt}
\rotate
\tablenum{6a}
\tabletypesize{\small}
\tablecaption{Abundances- HET/HRS\label{tab:absa}}
\tablehead{
     \colhead{}&
     \colhead{}&
     \colhead{HD\,20367}&
     \colhead{}&
     \colhead{HD\,40979}&
     \colhead{}&
     \colhead{HD\,52265H}&
     \colhead{}&
     \colhead{HD\,89744}&
     \colhead{}&
     \colhead{HD\,195019}&
     \colhead{}&
     \colhead{HD\,209458}&
     \colhead{}&
     \colhead{HD\,217107}
     }
\startdata
  
$\mathrm{[C/H]}$  && -0.01 $\pm0.05$   && +0.10 $\pm0.06$   && +0.14 $\pm0.05$  && +0.18 $\pm0.06$   && +0.06 $\pm0.03$  &&  -0.10 $\pm0.05$ && +0.32 $\pm0.07$ \\
$\mathrm{[N/H]}$  && $\leq$ +0.24      && $\leq$ +0.45	    && $\leq$ +0.30     && $\leq$ +0.42	     && +0.05 $\pm0.10$  && $\leq$ +0.15     && +0.49 $\pm0.09$ \\
$\mathrm{[O/H]}$  && +0.02 $\pm0.04$   && +0.16 $\pm0.06$   && +0.13 $\pm0.05$  && +0.22 $\pm0.05$   && +0.08 $\pm0.04$  && -0.04 $\pm0.05$  && +0.23 $\pm0.05$ \\
$\mathrm{[Na/H]}$ && +0.07 $\pm0.03$   && +0.27 $\pm0.04$   && +0.28 $\pm0.02$  && +0.27 $\pm0.03$   && +0.03 $\pm0.03$  && +0.02 $\pm0.04$  && +0.47 $\pm0.04$ \\
$\mathrm{[Mg/H]}$ && +0.08 $\pm0.02$   && +0.23 $\pm0.05$   && +0.22 $\pm0.05$  && +0.25 $\pm0.03$   && +0.11 $\pm0.03$  && +0.05 $\pm0.03$  && +0.41 $\pm0.04$ \\
$\mathrm{[Al/H]}$ && +0.03 $\pm0.02$   && +0.21 $\pm0.03$   && +0.21 $\pm0.02$  && +0.25 $\pm0.06$   && +0.07 $\pm0.02$  && +0.04 $\pm0.02$  && +0.44 $\pm0.03$ \\
$\mathrm{[Si/H]}$ && +0.10 $\pm0.01$   && +0.27 $\pm0.02$   && +0.24 $\pm0.02$  && +0.27 $\pm0.02$   && +0.07 $\pm0.01$  && +0.05 $\pm0.02$  && +0.39 $\pm0.02$ \\
$\mathrm{[S/H]}$  && +0.06 $\pm0.05$   && +0.16 $\pm0.05$   && +0.18 $\pm0.05$  && +0.17 $\pm0.04$   && +0.02 $\pm0.03$  &&\nodata	    && +0.40 $\pm0.09$ \\
\enddata

\end{deluxetable}

%% file: tab6b.tex
\begin{deluxetable}{lcllllllc}
\tablecolumns{9}
\tablewidth{0pt}
\rotate
\tablenum{6b}
\tabletypesize{\small}
\tablecaption{Abundances- ESO/FEROS\label{tab:absb}}
\tablehead{
     \colhead{}&
     \colhead{}&
     \colhead{HD\,2039}&
     \colhead{}&
     \colhead{HD\,55265F}&
     \colhead{}&
     \colhead{HD\,75289}&
     \colhead{}&
     \colhead{HD\,76700}
     }
\startdata
  
$\mathrm{[C/H]}$  && +0.25 $\pm0.06$   && +0.18 $\pm0.06$  && +0.08 $\pm0.05$  && +0.30 $\pm0.05$  \\
$\mathrm{[N/H]}$  && +0.41 $\pm0.09$   && $\leq$ +0.30     && $\leq$ +0.37     && +0.48 $\pm0.09$  \\
$\mathrm{[O/H]}$  && +0.22 $\pm0.06$   && +0.08 $\pm0.06$  && +0.04 $\pm0.05$  && +0.24 $\pm0.05$  \\
$\mathrm{[Na/H]}$ && +0.40 $\pm0.05$   && +0.28 $\pm0.04$  && +0.20 $\pm0.04$  && +0.40 $\pm0.06$  \\
$\mathrm{[Mg/H]}$ && +0.32 $\pm0.06$   && +0.17 $\pm0.06$  && +0.18 $\pm0.04$  &&\nodata	   \\
$\mathrm{[Al/H]}$ && +0.32 $\pm0.02$   && +0.22 $\pm0.02$  && +0.29 $\pm0.03$  && +0.48 $\pm0.02$  \\
$\mathrm{[Si/H]}$ && +0.34 $\pm0.02$   && +0.22 $\pm0.02$  && +0.25 $\pm0.02$  && +0.39 $\pm0.01$  \\
$\mathrm{[S/H]}$  && +0.31 $\pm0.06$   && +0.21 $\pm0.05$  && +0.08 $\pm0.04$  && +0.22 $\pm0.04$ \\
\enddata

\end{deluxetable}

%% file: tab7.tex
\begin{deluxetable}{lcccccccc}
\tablecolumns{9}
\tablewidth{0pt}
\tablenum{7}
\tablecaption{Abundance Sensitivities\label{tab:sens}}
\tablehead{
     \colhead{}&
     \colhead{}&
     \multicolumn{3}{c}{HD\,20367}&
     \colhead{}&
     \multicolumn{3}{c}{HD\,76700}\\
     \cline{3-5} \cline{7-9}\\
     \colhead{}&
     \colhead{}&
     \colhead{$\Delta T_{\rm eff}$}&
     \colhead{$\Delta \log g$}&
     \colhead{$\Delta \xi$}&
     \colhead{}&
     \colhead{$\Delta T_{\rm eff}$}&
     \colhead{$\Delta \log g$}&
     \colhead{$\Delta \xi$}\\
     \colhead{Species}&
     \colhead{}&
     \colhead{($\pm 150$ K)}&
     \colhead{($\pm 0.25$ dex)}&
     \colhead{($\pm 0.30 {\rm km s^{-1}}$)}&
     \colhead{}&
     \colhead{($\pm 150$ K)}&
     \colhead{($\pm 0.25$ dex)}&
     \colhead{($\pm 0.30 {\rm km s^{-1}}$)}     
     }
\startdata
\ion{Fe}{1} && $\pm$0.10 & $\mp$0.01 & $\mp$0.02 && $\pm$0.09 & $\mp$0.01 & $\mp$0.05\\
\ion{Fe}{2} && $\mp$0.04 & $\pm$0.11 & $\mp$0.03 && $\mp$0.07 & $\pm$0.10 & $\mp$0.06\\
\ion{C}{1}  && $\mp$0.09 & $\pm$0.08 & $\mp$0.01 && $\mp$0.10 & $\pm$0.06 & $\mp$0.01\\
\ion{N}{1}  && $\pm$0.22 & $\mp$0.01 & $\pm$0.01 && $\pm$0.24 & $\mp$0.01 & $\pm$0.01\\
\ion{O}{1}  && $\mp$0.13 & $\pm$0.07 & $\mp$0.03 && $\mp$0.16 & $\pm$0.04 & $\mp$0.04\\
(\ion{O}{1})&& $^{+0.01}_{-0.03}$ & $^{+0.03}_{-0.05}$ & $^{-0.01}_{-0.01}$ && $^{+0.02}_{-0.01}$ & $^{+0.11}_{-0.05}$ & $^{+0.01}_{+0.01}$\\
\ion{Na}{1} && $\pm$0.07 & $\mp$0.03 & $\mp$0.02 && $\pm$0.08 & $\mp$0.03 & $\mp$0.03\\
\ion{Mg}{1} && $\pm$0.07 & $\mp$0.03 & $\mp$0.03 && $\pm$0.06 & $\mp$0.05 & $\mp$0.10\\
\enddata

\end{deluxetable}

%% file: tab8.tex
\begin{deluxetable}{lcr}
\tablecolumns{3}
\tablewidth{0pt}
\tablenum{8}
\tablecaption{Condensation Temperature of the Elements\label{tab:condenTemp}}
\tablehead{
     \colhead{}&
     \colhead{}&
     \colhead{\tc\tablenotemark{a}}\\
     \colhead{Element}&
     \colhead{}&
     \colhead{(K)}
     }
\startdata
C   &&   40\\
N   &&  123\\
O   &&  180\\
Na  &&  958\\
Mg  && 1336\\
Al  && 1653\\
Si  && 1310\\
S   &&  664\\
Ca  && 1517\\
Sc  && 1659\\
Ti  && 1582\\
V   && 1429\\
Cr  && 1296\\
Mn  && 1158\\
Fe  && 1334\\
Co  && 1352\\
Ni  && 1353\\
Zn  &&  726\\
\enddata

\tablenotetext{a}{50\% condensation temperatures from \citet{2003ApJ...591.1220L}}

\end{deluxetable}

%% file: tab9.tex
\begin{deluxetable}{lrrrrcr}
\tablecolumns{7}
\tablewidth{0pt}
\tablenum{9}
\tablecaption{Abundance Slopes with \tc\label{tab:slopes}}
\tablehead{
     \colhead{}&
     \colhead{}&
     \colhead{}&
     \colhead{}&
     \colhead{}&
     \colhead{Slope}&
     \colhead{}\\
     \colhead{Star}&
     \colhead{}&
     \colhead{Slope\tablenotemark{a}}&
     \colhead{$\sigma$}&
     \colhead{}&
     \colhead{($T_{\mathrm{c}} > 900$ K)}&
     \colhead{$\sigma$}
     }
\startdata
HD\,20367  && 5.46  & 1.72  && -0.51  & 5.27 \\
HD\,40979  && 6.52  & 2.09  && -7.22  & 4.23 \\
HD\,52265H && 5.41  & 1.36  && -4.78  & 2.73 \\
HD\,89744  && 2.24  & 2.28  &&  1.55  & 7.10 \\
HD\,195019 && 0.81  & 1.54  &&  9.07  & 3.61 \\
HD\,209458 && 6.75  & 2.00  &&  5.73  & 6.00 \\
HD\,217107 && 4.73  & 3.13  && -9.06  & 5.70 \\
           &&       &       &&        &      \\
HD\,2039   && 4.88  & 1.85  && -9.56  & 3.46 \\
HD\,52265F && 5.19  & 2.22  && -9.37  & 4.20 \\
HD\,75289  && 14.26 & 2.08  &&  7.18  & 5.44 \\
HD\,76700  && 12.93 & 2.98  && 11.43  & 6.08 \\

\enddata

\tablenotetext{a}{all values are $\times 10^{-5}$ dex K$^{-1}$}

\end{deluxetable}

%% file: tab10.tex
\begin{deluxetable}{lccrcrc}
\tablecolumns{7}
\tablewidth{0pt}
\tablenum{10}
\tablecaption{Planet Properties\label{tab:planets}}
\tablehead{
     \colhead{}&
     \colhead{}&
     \colhead{$M_{\sin i}$}&
     \colhead{}&
     \colhead{Semi-Major Axis}&
     \colhead{Period}&
     \colhead{}\\
     \colhead{Star}&
     \colhead{}&
     \colhead{($M_{\mathrm{J}}$)}&
     \colhead{}&
     \colhead{(AU)}&
     \colhead{(days)}&
     \colhead{Eccentricity}
     }
\startdata
HD\,75289 b  && 0.5 && 0.05 &	 3.5 & 0.03\\
HD\,76700 b  && 0.2 && 0.05 &	 4.0 & 0.10\\ 
HD\,209458 b && 0.7 && 0.05 &	 3.5 & 0.00\\
HD\,195019 b && 3.6 && 0.14 &	18.2 & 0.01\\
HD\,52265 b  && 1.1 && 0.50 &  119.3 & 0.33\\
HD\,40979 b  && 4.0 && 0.85 &  264.1 & 0.25\\
HD\,89744 b  && 8.5 && 0.92 &  256.8 & 0.67\\
HD\,20367 b\tablenotemark{a}  && 1.1 && 1.25 &  500.0 & 0.23\\ 
HD\,2039  b  && 6.0 && 2.20 & 1120.0 & 0.72\\
HD\,217107 c && 2.6 && 5.33 & 4300.0 & 0.52\\
HD\,217107 b && 1.4 && 0.08 &	 7.1 & 0.13\\

\enddata

\tablenotetext{a}{Data taken from the Extrasolar Planet Encyclopedia 
(available at http://exoplanet.eu).  For all other stars, data taken from the
Exoplanet Data Explorer (available at http://exoplanets.org).}

\end{deluxetable}